\documentstyle[12pt,epsf]{article}
\begin{document}
\thispagestyle{empty}

\vspace*{1.5cm}

\begin{center}
{\large \bf
Elastic Proton-Deuteron Backward Scattering:\\
Relativistic Effects and Polarization Observables
}\\[1cm]

{\sc
L.P. Kaptari$^{a,b}$,
B. K\"ampfer$^a,$
S.M. Dorkin$^d$,
S.S. Semikh$^b$}\\[1cm]

$^a$Research Center Rossendorf, Institute for Nuclear and Hadron Physics,\\
PF 510119, 01314 Dresden, Germany\\[1mm]
$^b$Bogoliubov Laboratory of Theoretical Physics, JINR Dubna,\\
P.O. Box 79, Moscow, Russia \\[1mm]
$^d$ Far-Eastern State University, Vladivostok, Russia
\end{center}

\vskip 1cm

\begin{abstract} 
The  elastic proton-deuteron backward reaction is analyzed
within a covariant approach based on the Bethe-Salpeter
equation with realistic meson-exchange interaction.
Lorentz boost and other relativistic effects in the cross section
and spin correlation observables, like tensor analyzing power and
polarization transfer etc.,
are investigated in explicit form. Results of numerical calculations for
a complete set of polarization observables are presented.
\end{abstract}

\vspace*{1cm}

\noindent
key words: elastic proton-deuteron scattering,
Bethe-Salpeter equation,\\
\phantom{key words:}
polarization observables\\[3mm]
PACS number(s): 21.45+v, 21.10.Ky, 21.60.-n

\newpage
\section{Introduction} 

Presently a wideranging program of precisely investigating the structure of
the lightest nuclei is under consideration.
There are new proposals to study the polarization
characteristics of the deuteron using both hadronic \cite{proposal,cosy}
and electromagnetic probes (cf.~\cite{cebaf,arenh}).
Besides the goal of checking fundamental results
of quantum chromodynamics (for instance, the study of
the $Q^2$ evolution of the Gerasimov-Drell-Hearn sum rule \cite{cebaf})
the envisaged experiments focus on a complete reconstruction of the
amplitude of the corresponding process \cite{proposal,arenh,rekalo} and
an overall investigation of the nuclear momentum distribution
\cite{cosy,experiment}.

The simplest reactions with hadron probes are
processes of forward or backward scattering of protons
off the deuteron. The extensive experimental study of these
reactions has started a decade ago in Dubna
and Saclay (cf. \cite{experiment,dpelastic,punj95,belost})
and is planned to be continued in the nearest future at COSY \cite{cosy}.
One may classify this type of reactions as inclusive
break-up processes, exclusive quasi-elastic
scattering and elastic processes. A common feature of
these processes is that in a collinear geometry,
the measured momenta of the fragments
are directly connected with the argument of the deuteron
wave function in the momentum space,
supposed the reaction mechanism is dominated
by the one-nucleon exchange. In such a way
a direct experimental investigation of the momentum distribution
within the deuteron in a large interval of internal momenta seems
to be accessible.
By using  polarized particles one may investigate as well
different aspects of spin-orbit interaction in the deuteron
and obtain hints on the role of
non-nucleon degrees of freedom in the deuteron wave function,
like $\Delta$ isobars, $N\bar N$ excitations and so on.
An encouraging fact here is that the extracted momentum distributions
from different reactions with electromagnetic and hadron probes
are rather similar, and therefore a realization of experimental programs
at different facilities may provide a quite complete
information on the internal structure of the deuteron.

Nowadays  the elastic proton-deuteron ($pD$)
backward scattering with both polarized protons and deuterons
receives a renewed interest \cite{cosy}.
A distinguished peculiarity of this process is that
within the impulse approximation the cross section is
proportional to the fourth power of the deuteron wave
function, contrary to the break-up and quasi-elastic
reactions which are proportional to the second power of the wave function.
This makes the processes of elastic scattering much more
sensitive to the theoretically assumed mechanisms, and
even a slight modification of the deuteron wave function
may result in significant deviations from the
calculated cross section.
However, as pointed out by Vasan \cite{wassan},
Frankfurt and Strikman \cite{strik} and Karmanov \cite{karm}
the polarization observables, like tensor analyzing
power and polarization transfer, are exactly
as those obtained for the break-up and quasi-elastic scattering
in the non-relativistic limit.
Beyond both the non-relativistic
limit and the impulse approximation the polarization observables
differ for different processes. Hence, a combined analysis
of data on polarization characteristics from the above mentioned
three processes will constrain the basic reaction mechanism and the role
of non-nucleon degrees of freedom and relativistic
effects in the deuteron. Another
peculiarity of elastic backward or forward $pD$
reactions is that the amplitude of the processes is
determined by only four complex helicity amplitudes,
and a complete reconstruction
of these amplitudes seems possible in one experimental set-up.
For this it is sufficient to measure 10 independent observables
as proposed in refs.~\cite{rekalo,ladygin}.
Certainly this does not mean that the realization
of such an exhausting experiment
will determine entirely the deuteron structure;
only within the non-relativistic impulse approximation
the cross section is directly related to the deuteron wave function.

First measurements of polarization observables, such as the
tensor analyzing power $T_{20}$ and polarization transfer
$\kappa$, have been performed in Dubna~\cite{dpelastic,ls,azh97}
and Saclay \cite{punj95,belost}.
Theoretically the
elastic $pD$ scattering has been studied by many authors
\cite{kislind,keisterTj,keister,wilkin,nakamura}. It has been shown that
the cross section can not be satisfactorily described within
the non-relativistic impulse approximation and that other
mechanisms, e.g., described by meson-exchange triangle
diagrams \cite{wilkin,nakamura}, are important.
Besides the importance of other mechanisms, the role
of relativistic corrections within the impulse approximation
has been studied by several authors already some time ago
(see \cite{keisterTj,keister} and further references therein) within the
Bethe-Salpeter (BS) formalism. The unpolarized   cross
section and the tensor analyzing power have been
numerically computed in a
fully covariant way and a comparison with
polarization data, available at this time, has been made. However, in view of
the present experimental situation and future
proposals \cite{proposal,cosy,rekalo,ladygin}
a detailed covariant investigation of the role of relativistic corrections,
such as Lorentz boost effects and contributions of negative-energy waves etc.\
is still lacking. In the present paper an attempt
is presented to fill this gap.

We focus here on a detailed study of the elastic
$pD$ amplitude within the BS approach by using the numerical solution
obtained with a realistic one-boson exchange
interaction \cite{parametrization,solution}. It is known
that one of the unpleasant features within the BS formalism
is the cumbersomeness of the final expressions for the calculated
observables  and  difficulties in their physical interpretation
(cf.\ refs.~\cite{keisterTj,gross,quad}).
In this paper we try to avoid this problem and
present our results in a form as simple as possible.
For this sake we separate the contributions of the positive-energy waves
and identify them in the non-relativistic limit.
The contributions of the Lorentz boost effects and the
negative-energy waves in the deuteron are then calculated in leading order.
That means, in the fully covariant results we keep the first
orders of negative-energy $P$ waves and the leading-order terms
of a Taylor expansion. In this way
we are able to separate and to investigate in explicit
form the contribution of the Lorentz boost effects and
negative-energy waves to the amplitude, cross section and
polarization observables as well. Results of numerical calculations
for a complete set of observables are also presented.

Our paper is organized as follows:
In section II the kinematics of the process is
described, and the covariant amplitude is derived in details in section III.
Since the covariant expressions within the BS approach are
rather lengthy and since in the procedure of the arrangement of
results in parts containing non-relativistic formulae and boost
effects and relativistic corrections separately, it is very important
to provide a clear definition of all the relevant variables.
In sections IV - VI a complete set of polarization observables is
defined in terms of BS wave functions. The non-relativistic
limit, Lorentz boost effects and relativistic corrections
are studied analytically and numerically for the cross section
and for a class of polarization observables. A comparison with
available experimental data is also made.
In section VII an interpretation of the relativistic corrections
in terms of non-relativistic meson-exchange like contributions
is performed. To do so we solve the BS equation for the negative-energy
$P$ waves in the one-iteration approximation and express
explicitly the $^3P_1^{+-}$ and
$^1P_1^{+-}$ waves via the non-relativistic $S$ and $D$ waves of the
deuteron. The obtained result for the elastic amplitude
is found in a form being very similar to amplitudes
computed in the non-relativistic picture when estimating the role of
$N\bar N$ pair currents in electromagnetic processes.
Some cumbersome expressions and useful formulae
are collected in the Appendices A and B.

\section{Kinematics} 

We consider the elastic backward scattering reaction of the type
\begin{equation}
 p \, + \ D = p'(\theta = 180^0)\, + \, D'.
\label{reaction}
\end{equation}
The differential cross section of the reaction (\ref{reaction})
in the center of mass system (c.m.s.) of colliding particles reads
\begin{equation}
\frac{d\sigma}{d\Omega}\,=\,\frac{1}{64\pi^2s}\,
|{\cal M}\,|^2,
\label{cross}
\end{equation}
where $s$ is the Mandelstam variable
denoting the total energy squared in the c.m.s., and
${\cal M}$ is the invariant amplitude of the process. In the
case of backward scattering the cross section eq.~(\ref{cross}) depends only
on one kinematical variable which usually is chosen as
$s$. Other variables can be expressed via $s$ by using
energy conservation. For instance, the Mandelstam variable $u$
is $u=(M_d^2-m^2)^2/s$, while the c.m.s. momentum is $\mbox{\boldmath{$p$}}\, ^{2}=-t/4$ etc.
Here $M_d$ and $m$ stand for the deuteron and nucleon masses, respectively.
Then in the c.m.s
we define the relevant kinematical variables as follows:
the four-momenta of particles read
\begin{equation}
D=(E,\mbox{\boldmath{$p$}}),\,\,p=(\epsilon,-\mbox{\boldmath{$p$}}),\,\,
D'=(E,-\mbox{\boldmath{$p$}}),\,\,p'=(\epsilon,\mbox{\boldmath{$p$}}),
\label{moms}
\end{equation}
and the polarization four-vectors of the deuteron with polarization
indices $M$ and $M'$ can be written as
\begin{eqnarray}
&&
\xi_M=(\frac {\mbox{\boldmath{$p$}} \mbox{\boldmath{$\xi$}}_M}{M_d}, \mbox{\boldmath{$\xi$}}_M+\mbox{\boldmath{$p$}} \frac {\mbox{\boldmath{$p$}} \mbox{\boldmath{$\xi$}}_M}{M_d(E+M_d)})
\label{xi}\\[2mm]
&&
\xi'_{M'}=(-\frac{\mbox{\boldmath{$p$}} \mbox{\boldmath{$\xi$}}'_{M'}}{M_d},
\mbox{\boldmath{$\xi$}}'_{M'}+\mbox{\boldmath{$p$}} \frac {\mbox{\boldmath{$p$}} \mbox{\boldmath{$\xi$}}'_{M'}}{M_d(E+M_d)}),
\label{xi1}
\end{eqnarray}
where $\mbox{\boldmath{$\xi$}} ,\mbox{\boldmath{$\xi$}}'$ are the three-polarization vectors
in the  rest frame of the deuteron,
\begin{equation}
\mbox{\boldmath{$\xi$}}_{+ 1} =(-1,i,0)/\sqrt{2},\quad \mbox{\boldmath{$\xi$}}_{- 1} =(1,i,0)/\sqrt{2},
\quad
\mbox{\boldmath{$\xi$}}_0=(0,0,1).
\label{xilab}
\end{equation}
The proton spinors are normalized as $\bar{u}(p) u(p)=2m$ with
\begin{eqnarray}
u(\mbox{\boldmath{$p$}},s)=\sqrt{m+\epsilon}
\left ( \begin{array}{c} \chi_s \\ -\frac{\mbox{\boldmath{$\sigma$}} \mbox{\boldmath{$p$}}}{m+\epsilon}\chi_s
\end{array}  \right) \quad \quad
u(\mbox{\boldmath{$p$}}\,',s')=\sqrt{m+\epsilon}\left( \begin{array}{c} \chi_{s'} \\
\frac{\mbox{\boldmath{$\sigma$}} \mbox{\boldmath{$p$}}}{m+\epsilon}\chi_{s'}
\end{array}  \right),
\label{spinors}
\end{eqnarray}
where $\chi_s$ denotes the usual two-dimensional Pauli spinor.

In what follows we shall widely exploit as variable the momentum of the
outgoing proton, $P_{lab}$, in the rest frame of the incoming deuteron,
which we here define as laboratory system.
The relation between c.m.s. and
laboratory system is simply expressed by
$ |\mbox{\boldmath{$p$}}|\, = 2mP_{lab}/\sqrt{u}$.

Now we proceed with an  analysis of the general properties
of the invariant amplitude ${\cal M}$. In principle,
the amplitude  ${\cal M}$ for the elastic fermion-vector--boson scattering
has been  studied in detail and is well known
(see, for instance refs.~\cite{rekalo,ladygin,ghazik}),
nevertheless for the sake of completeness we present here
some of the most important characteristics of ${\cal M}$.

The process of the elastic $pD$ scattering is
determined by 12 independent partial amplitudes \cite{ghazik}.
However, in case of forward
or backward scattering, due to the conservation of the total helicity
of colliding particles, only four
amplitudes remain independent, and these four
amplitudes determine all the possible polarization observables
of the process. There are many ways of representing these
four amplitudes. In order to emphasize explicitly the
transition between initial and final states
with fixed helicities it is convenient to represent
${\cal M}$ in the c.m.s.\ in a two-dimensional spin space for the
proton spinors and three-dimensional space for the
deuteron spin characteristics.
In this case the manifest covariance of the amplitude is lost.
However the analysis and final formulae become much simpler
and transparent. Moreover, by making use of
eqs.~(\ref{moms}) - (\ref{spinors})
all the polarization characteristics of the reaction may be
expressed via the corresponding quantities
evaluated in the deuteron rest frame. This gives another advantage of
such an analysis, namely it allows for a straightforward non-relativistic
limit, in particular avoiding the problem of boosting or not the
non-relativistic polarization vectors from the rest frame to the
c.m.s. It is worth emphasizing that
in such a representation of the invariant amplitude,
in spite of the fact that it is not explicitly
covariant, its form is the most  general one
and valid in both the relativistic picture
and the non-relativistic limit as well.

In this paper we keep our notation as close as possible to the one
used in refs.~\cite{rekalo,ladygin}. Hence the total
amplitude is written in the form
\begin{equation}
{\cal M} =  \chi_{s'}^+\, {\cal F} \,\chi_s
\label{ampF}
\end{equation}
with
\begin{eqnarray}
{\cal F} = {\cal A}\,(\mbox{\boldmath{$\xi$}}_M\mbox{\boldmath{$\xi$}}^+_{M'})+
{\cal B}\,(\mbox{\boldmath{$n$}}\mbox{\boldmath{$\xi$}}_M)(\mbox{\boldmath{$n$}}\mbox{\boldmath{$\xi$}}^+_{M'})
+i{\cal C}\,(\mbox{\boldmath{$\sigma$}}\cdot [\mbox{\boldmath{$\xi$}}_M\times\mbox{\boldmath{$\xi$}}^+_{M'}])
+i{\cal D}\,(\mbox{\boldmath{$\sigma$}}\mbox{\boldmath{$n$}})(\mbox{\boldmath{$n$}}\cdot [\mbox{\boldmath{$\xi$}}_M\times\mbox{\boldmath{$\xi$}}^+_{M'}]) ,
\label{amplit}
\end{eqnarray}
where $\mbox{\boldmath{$n$}}
$ is a unit vector parallel to the beam direction;
${\cal A,B,C}$ and ${\cal D}$ are the partial amplitudes of the $pD$ elastic
scattering process depending on the initial energy.
Then the cross section (\ref{cross}) is determined
by Tr$({\cal F}^+\,{\cal F})$. In the following we suppress the subscripts
$M$ and $M'$ of the polarization vectors $\xi$,
bearing in mind that in computing observables
the summation over these indices results in the
completeness relation for $\xi$ or in the polarization
density matrix of the deuteron, which
reads in covariant form \cite{bayer}
\begin{eqnarray}
&&
\sum_{M}\,\xi_M^\mu\xi_M^{+\nu} \, = \,
\left ( -g_{\mu\nu}  + \frac{D^\mu D^\nu}{M_d^2}\right),
\label{complet}\\[2mm]
\rho_{\mu\nu} &=& \frac{1}{3}\left (-g_{\mu\nu}+
\frac{D_{\mu}D_{\nu}}{M_d^2}\right )
+\frac{1}{2M_d}i
{\epsilon}_{\mu\nu\gamma\delta} D^{\gamma} {\cal S}_{D}^{\delta}
\nonumber\\
&&+\left \{-\frac{1}{2}
\left(
{(W_{\lambda_1})}_{\mu\rho}
{(W_{\lambda_2})}^{\rho}_{~\nu}+
{(W_{\lambda_2})}_{\mu\rho}
{(W_{\lambda_1})}^{\rho}_{~\nu}
\right) \right.
\label{rhocovariant}\\
&& \left. - \frac{2}{3}
 \left (-g_{\lambda_1\lambda_2}+\frac{D_{\lambda_1}D_{\lambda_2}}{M_d^2}\right )
\left (-g_{\mu\nu}+\frac{D_{\mu}D_{\nu}}{M_d^2}\right) \right\}
Q_{D}^{\lambda_1
\lambda_2},
\nonumber\end{eqnarray}
where
${(W_{\lambda})}_{\mu\nu} \equiv
i \epsilon_{\mu\nu\gamma\lambda}D^{\gamma}/M_d$;
${\cal S}_{D}$ is the spin vector, and $Q_{D}$ stands for
the alignment tensor of the deuteron.
$\mu, \nu, \lambda \cdots$ are Lorentz indices, and
we use the metric $g_{\mu \nu}$ with signature $-2$.

In the three-dimensional representation for the deuteron
polarization and two-dimensional Pauli matrices for the nucleon
states the corresponding density matrices can be cast in the simple form
\begin{eqnarray}
&&
\rho_p = \frac{1}{2}\left ( I + (\mbox{\boldmath{$\sigma$}}\, {\bf P}_p)
\right ),
\label{denprot}\\[2mm]
&&
\rho^{\alpha\beta}=\frac 13
\left( \delta_{\alpha\beta} -\frac 32 i\epsilon^{\alpha\beta\delta}
\tilde {\cal S}^\delta_D - 2\,\tilde Q_D^{\alpha\beta}\right ),
\label{dendeut}
\end{eqnarray}
where ${\bf P}_p$ is the proton polarization three-vector,
$\tilde {\cal S}_D$ and $ \tilde Q_D$ are spin and tensor
polarization operators (actually  $3\times3$ matrices) of the deuteron.

\section{The one-nucleon exchange mechanism} 

We investigate here the relativistic one-nucleon exchange
defined by the diagram depicted in fig.~1.
Using the kinematics shown in fig.~1
and working with the Mandelstam technique \cite{Mandelstam_tech},
the one-nucleon exchange contribution to the elastic amplitude within the
BS formalism is
\begin{eqnarray}
{\cal M} = \bar u(p') \, \Gamma (D,q) \, \tilde{S_2} \,
\bar{\Gamma} (D',q') \, u(p).
\label{gn}
\end{eqnarray}
$\Gamma(D,q)$ denotes the BS vertex function of the deuteron;
$ \tilde{S_2} = 1/\left (\hat D/2-\hat q+m \right)$
is the nucleon propagator,
and $\bar{\Gamma}=\gamma_0 \Gamma^+ \gamma_0$.
We use the abbreviation $\hat D = D^\mu \gamma_\mu$
when contacting a four-vector with Dirac matrices.
The momenta $q$ and $q'$ are fixed by the conditions $D/2+q=p'$
and $D'/2+q'=p$.
The vertex function $\Gamma(D,q)$ is
the solution of the BS equation for the deuteron bound state.
The BS equation and consequently its solution
$\Gamma(D,q)$ are sixteen-component objects in the spinor space.
To solve the BS equation and to compute observables within the BS
formalism one usually represents the vertex function $\Gamma(D,q)$
as a $4\times 4$ matrix and utilizes a decomposition of  $\Gamma(D,q)$
over a complete set of matrices in the sixteen-dimensional
spinor space. As mentioned in ref.~\cite{quad}, the choice of the
representation of the matrices depends on the special attacked problem.
Actually in some calculations it is convenient to combine two representations,
namely the complete set of Dirac matrices (to perform explicit numerical
calculations) and the so-called $\rho$ spin classification of the partial
vertices~\cite{cubis}
(to have a more transparent physical interpretation
of the obtained results; for details consult ref.~\cite{quad,semikh}).
In the present paper we use mainly the $\rho$ spin classification,
however the numerical calculations have been performed in terms
of the solution of the BS equation obtained in the Dirac
basis \cite{parametrization,solution,khanna}.
Notice that the solution of the BS equation
has been obtained in the deuteron rest frame, whereas considering
the amplitude (\ref{gn}) it is seen that at least one
vertex $\Gamma(D,q)$ is defined in a system where the deuteron
is moving.
Obviously in this case one may explicitly boost the
vertex $\Gamma(D,q)$ from the rest frame to the c.m.s.\
and analyze the amplitude (\ref{gn}) in terms
of a BS solution at rest and Lorentz boost effects separately.
Our experience shows that in this way one obtains rather cumbersome
expressions and a relatively simple and  physically meaningful
analysis of results is straitened.
Therefore, we represent the BS vertex function in a covariant form
with eight invariant scalar functions (see Appendix A).
Observing that in the process depicted in fig.~1,
in each vertex  $\Gamma(D,q)$ one nucleon is on the mass shell,
the  part contributing to the amplitude (\ref{gn})  may be
written in a form exactly coinciding with the one used,
for instance, by Gross \cite{gross} or
Keister and Tjon \cite{keisterTj,bselastic}:
\begin{eqnarray}
\Gamma(D,q) & = &
[h_1 \hat{\xi} +h_2 \frac {(q \xi)}{m}] +
[h_5 \hat {\xi} +h_6 \frac {(q \xi)}{m}] \frac {\hat{D}/2-\hat{q}+m}{m},
\label{vf}
\end{eqnarray}
where $h_i$ are invariant scalar functions depending on the
invariants  $q^2$ and $Dq$.
Then after summation over the spins in eq.~(\ref{cross})
one gets the cross section in a fully covariant form
in terms of the relativistic BS solutions $\Gamma(D,q)$ as
\begin{eqnarray}
\frac{d\sigma}{d\Omega} & = & \frac{1}{64\pi^2s}\,
\frac 16 \mbox{Tr}
\left [ (\hat p\,'+m)
\Gamma (D,q) \tilde{S_2} \bar{\Gamma} (D',q')
(\hat p\,+m) \Gamma (D',q')\tilde{S_2}
\bar\Gamma (D,q) \right],
\label{crossBS}
\end{eqnarray}
where, when replacing $\Gamma(D,q)$ by  eq.~(\ref{vf})
and summing over the deuteron spins,
one should make use of eqs.~(\ref{complet}) or
(\ref{rhocovariant})
in dependence  of the initial and final states
of the particles in a concrete measurement.

Eq.~(\ref{crossBS}) together with eqs.~(\ref{complet})
and  (\ref{rhocovariant}) completely determine
all characteristics of the process.
When computing the trace in eq.~(\ref{crossBS})
one obtains a fully covariant relativistic
expression of all observables. In our calculations we use
a suitable algebraic formula manipulation code which,
within the representation of the solution of the BS equation in the
form (\ref{vf}),
delivers the covariant, but rather cumbersome results
(for examples cf.\ ref.~\cite{keisterTj}). These
results have been tested by evaluating  the non-relativistic limits
for the cross section and polarization observables. However,
further investigations relying on these lengthy analytical expressions
seem to be almost impossible.
Therefore, for an explicit study of the influence of
the Lorentz boost effects and other relativistic
corrections we shall investigate different aspects of
the amplitude ${\cal M}$ in eq.~(\ref{cross})
instead of the cross section (\ref{crossBS}).

By substituting eq.~(\ref{vf}) into eq.~(\ref{gn})
and making use of the Gordon identity,\\
$\bar u(p',s') [\hat a(\hat p-m) +(\hat p' -m)\hat a] u(p,s)=0$,
one can represent the amplitude in the form
\begin{equation}
{\cal M} =\sum_{i=1}^6\tilde R_i \, \bar u(p') \, R_i \, u(p)
\label{amplit1}
\end{equation}
with six invariant scalar functions $\tilde R_i$
and six covariant spin structures defined as
\begin{eqnarray}
&&
R_1 =\hat{\xi} \hat {\xi}', \quad
R_2=\frac {p \xi'}{m}\frac {p' \xi}{m}, \quad
R_3=\hat{\xi} \frac {p \xi'}{m}+
\hat{\xi}' \frac {p \xi'}{m},
\quad
R_4=\hat{\xi} (\hat{D}-\hat{p}') \hat {\xi}',
\label{r14}
\\[2mm]
&&
R_5=\frac {p \xi'}{m} \frac {p' \xi}{m}
(\hat{D}-\hat{p}'),
\label{r45} \quad
R_6= \hat{\xi} (\hat{D}-\hat{p}') \frac {p \xi'}{m}+
(\hat{D}-\hat{p}') \hat {\xi}' \frac {p' \xi}{m},
\label{r56}
\end{eqnarray}
\begin{eqnarray}
&&
\tilde{R}_1 =\frac{1}{m}(2h_1h_5^*+h_5h_5^*) -
\frac {mh_1h_1^*}{(D-p')^2-m^2},
\label{tilder1}\\[2mm]
&&
\tilde{R}_2=
\frac{1}{m}(h_6h_6^*+2h_2h_6^*)- \frac {mh_2h_2^*}{(D-p')^2-m^2},
\label{tilder2}\\[2mm]
&&
\tilde{R}_3=\frac{1}{m}(h_1h_6^*+h_2h_5^*+h_5h_6^*)-
\frac {mh_1h_2^*}{(D-p')^2-m^2},
\label{tilder3}\\[2mm]
&&
\tilde{R}_4=\frac {h_1h_1^*}{(D-p')^2-m^2}+\frac {h_5h_5^*}{m^2},
\label{tilder4}\\[2mm]
&&
\tilde{R}_5= \frac {h_2h_2^*}{(D-p')^2-m^2} +\frac {h_6h_6^*}{m^2},
\label{tilder5}\\[2mm]
&&
\tilde{R}_6 =\frac {h_1h_2^*}{(D-p')^2-m^2} +\frac {h_5h_6^*}{m^2}.
\label{tilder6}
\end{eqnarray}
At first glance there seems to be a contradiction between eqs.~(\ref{r14}) and
(\ref{r56}) and the general expression (\ref{amplit}), namely
instead of four amplitudes our result contains six different
structures. However it is straightforward to prove
that the six covariant spin structures in eqs.~(\ref{r14}) and
(\ref{r56}) in the collinear kinematics reduce to exactly four independent
forms. For instance, taking into account that, in case of
forward or backward
elastic scattering, the expression $(\hat D-\hat p')$ has no spatial
component in the c.m.s.\ the three structures
$R_2, R_5$  and $R_6$ are equivalent and determine
the amplitude ${\cal B}$ in eq.~(\ref{amplit})
(see below). The structure $\hat \xi\, \hat \xi'$ may be cast into
the form of eq.~(\ref{amplit}) by exploiting Dirac's matrix algebra,
\begin{eqnarray}
&&
\hat \xi\, \hat \xi'= (\xi \xi')\,-\,i \xi_\mu\xi'_\nu
\,\sigma^{\mu\nu}= (\xi \xi')\,+\gamma_0
(\vec {\cal P}_1\vec \gamma) +i\gamma_0\gamma_5
(\vec {\cal P}_2\vec \gamma),
\label{hats}\\[2mm]
&&
\vec {\cal P}_1\equiv (\xi_0\vec\xi\,' + \xi'_0\vec \xi), \quad
\vec {\cal P}_2^\rho \equiv \epsilon_{\alpha\beta\rho}\xi_\alpha\xi'_\beta;
\quad \alpha,\beta,\rho=1,2,3.
\label{str}
\end{eqnarray}
Then it is seen that among the six amplitudes eqs.~(\ref{r14}) and
(\ref{r56}) only four are independent in collinear kinematics.
The correspondence between eq.~(\ref{amplit}) and eqs.~(\ref{r14}) and
(\ref{r56}) becomes obvious if the former are written in the c.m.s.
In this case,
\begin{eqnarray}
\bar u(p',s')\, R_1 \, u(p,s) & = &
\chi^+_{s'} \left[ -(\mbox{\boldmath{$\xi$}} \mbox{\boldmath{$\xi$}}') 2 \epsilon
-i \mbox{\boldmath{$\sigma$}}( \mbox{\boldmath{$\xi$}} \times \mbox{\boldmath{$\xi$}}') \frac EM_d 2m \right.
\label{r1cm} \\
& + & \left. i (\mbox{\boldmath{$\sigma$}} \mbox{\boldmath{$p$}}) ( \mbox{\boldmath{$p$}}, \mbox{\boldmath{$\xi$}} \times \mbox{\boldmath{$\xi$}}')
\frac{(m+\epsilon)^2-(M_d+E)^2}{M_d(M_d+E)(m+\epsilon)}
+
\frac{\mbox{\boldmath{$p$}} \mbox{\boldmath{$\xi$}}}{M_d} \frac {\mbox{\boldmath{$p$}} \mbox{\boldmath{$\xi$}}'}{M_d}(4E-4\epsilon)\right ]\chi_s,
\nonumber
\\[2mm]
\bar u(p',s')\,R_2\, u(p,s) & = & -
\frac {2\epsilon (E-\epsilon)^2}{ m^2}\chi^+_{s'}  \left[
\frac {\mbox{\boldmath{$p$}} \mbox{\boldmath{$\xi$}}}{M_d} \frac {\mbox{\boldmath{$p$}} \mbox{\boldmath{$\xi$}}'}{M_d}\right ]\chi_s,
\label{r2cm} \\
\bar u(p',s')\,R_3\, u(p,s) & = &
\frac {E-\epsilon}{M_dm}
\label{r3cm} \\
& \times &
\chi^+_{s'} \left [4mM_d \frac {\mbox{\boldmath{$p$}} \mbox{\boldmath{$\xi$}}}{M_d} \frac {\mbox{\boldmath{$p$}} \mbox{\boldmath{$\xi$}}'}{M_d}
+2 i \mbox{\boldmath{$\sigma$}}( \mbox{\boldmath{$\xi$}} \times \mbox{\boldmath{$\xi$}}') p^2
-2 i (\mbox{\boldmath{$\sigma$}} \mbox{\boldmath{$p$}}) ( \mbox{\boldmath{$p$}}, \mbox{\boldmath{$\xi$}} \times \mbox{\boldmath{$\xi$}}')\right ] \chi_s,
\nonumber \\[2mm]
\bar u(p',s')\,R_4\, u(p,s) & = &
-(E-\epsilon) \chi^+_{s'}
\left[ \phantom{ \frac{2(\epsilon^2-m^2-E\epsilon)}{M_d}}
-(\mbox{\boldmath{$\xi$}} \mbox{\boldmath{$\xi$}}')2m
+i (\mbox{\boldmath{$\sigma$}} \mbox{\boldmath{$p$}}) ( \mbox{\boldmath{$p$}}, \mbox{\boldmath{$\xi$}} \times \mbox{\boldmath{$\xi$}}')
\right.
\nonumber \\
& \times &
\left.
\frac {(M_d+E-m-\epsilon)^2}{M_d(M_d+E)(m+\epsilon)}
+i \mbox{\boldmath{$\sigma$}}( \mbox{\boldmath{$\xi$}} \times
\mbox{\boldmath{$\xi$}}') \frac{2(\epsilon^2-m^2-E\epsilon)}{M_d}
\right] \chi_s,
\label{r4cm} \\[2mm]
\bar u(p',s')\,R_5\, u(p,s) & = & -\frac{2(E-\epsilon)^3}{m}
\chi^+_{s'}
\left [\frac {\mbox{\boldmath{$p$}} \mbox{\boldmath{$\xi$}}}{M_d} \frac {\mbox{\boldmath{$p$}} \mbox{\boldmath{$\xi$}}'}{M_d} \right ] \chi_s,
\label{r5cm} \\[2mm]
\bar u(p',s')\,R_6\, u(p,s) & = & -\frac{4(E-\epsilon)^3}{m} \chi^+_{s'}
\left [\frac {\mbox{\boldmath{$p$}} \mbox{\boldmath{$\xi$}}}{M_d} \frac {\mbox{\boldmath{$p$}} \mbox{\boldmath{$\xi$}}'}{M_d}  \right ] \chi_s.
\label{r6cm}
\end{eqnarray}
It is worth stressing again that, in spite of these six covariant
spin structures $R_1 \cdots R_6$ being written in c.m.s.\
have lost their explicit covariance, they still determine
the covariant amplitude
${\cal M}$ of the process. The corresponding invariant scalar functions
$\tilde R_i$ defined by eqs.~(\ref{tilder1}) - (\ref{tilder6})
may be computed in any reference frame. Since the numerical
solutions~\cite{solution} of the BS equation have been obtained
in the deuteron rest frame, we also express $\tilde R_i$ in this system.
When one nucleon is on the mass shell, i.e. $M_d/2+q_0=E_p^{'}$,
the invariant functions $h_i$ are of the form (see Appendix A)
\begin{eqnarray}
&&
\sqrt{4\pi}h_1=\frac{1}{\sqrt{2}}g_1-\frac{1}{2}g_3 +
\frac{\sqrt{3}m}{2M_dP_{lab}}(2E_p^{'}-M_d)g_5,
\label{h16_1} \\&&
\sqrt{4\pi}h_2=-\frac{m}{\sqrt{2}(m+E_p^{'})}g_1-
\frac{m(m+2E_p^{'})}{2P_{lab}^2}g_3 +
\frac{\sqrt{3}m}{2P_{lab}M_d}(2M_d-E_p^{'})g_5,
\\&&
\sqrt{4\pi}h_5=-\frac{\sqrt{3}mE_p^{'}}{2M_dP_{lab}}g_5,
\\&&
\sqrt{4\pi}h_6=-\frac{m^2}{\sqrt{2}M_d(m+E_p^{'})}g_1+
\frac{(E_p^{'}+2m)m^2}{2M_dP_{lab}^2}g_3 +
\frac{\sqrt{6}m^2}{2M_dP_{lab}}g_7,
\label{h16}
\end{eqnarray}
where $g_i$ are the BS vertex functions in the
deuteron rest system and all the kinematical variables in
eqs.~(\ref{h16_1}) - (\ref{h16}) should be evaluated in this system.

\section{Observables} 

Having determined the amplitude by
eqs.~(\ref{amplit1}) - (\ref{h16}) one may define various
polarization characteristics of the process.
Employing the notation used in refs. \cite{rekalo,ladygin,ghazik}
we define the set of all possible polarization observables for the
non-covariant amplitude (\ref{ampF}) by
\begin{equation}
{\cal H}_{\lambda,H\to \lambda',H'} = \frac{\mbox{Tr} \left
({\cal F}\sigma_\lambda{\cal D}_H{\cal F}^+\sigma_{\lambda'}
{\cal D}_{H'} \right )}{\mbox{Tr} \left ({\cal F}{\cal F}^+\right )},
\label{observables}
\end{equation}
where the subscripts $\lambda$ and $H$
($\lambda'$ and $H'$) refer to the polarization
characteristics of the initial (final) proton and deuteron respectively;
$\sigma_\lambda$ is the Pauli matrix,
and ${\cal D}$ stands for a set of $3 \times 3$
operators defining the deuteron polarization. Note that
the introduced subscripts may appear as either single index or double indices
in dependence on the reaction conditions. For instance,
$0,0\to 0,0$ means a process with  unpolarized particles,
while $0,NN \to 0,NN$ means the tensor-tensor polarization
of the initial and final deuterons parallel
to the normal of the reaction plane direction.

At this point it is worth mentioning that
the numerical solution of the BS equation has been obtained
in the Euclidean space-time with imaginary time component $q_0$ of
the relative momentum $q$.
In  the process under consideration $q_0$ is fixed and real.
Hence, one needs either a numerical procedure for an analytical
continuation of the amplitudes to the real relative
energy axis (cf. \cite{keisterTj}) or another recipe \cite{anatomia}
for using the numerical solutions in this case.

We rely on the analysis  of the BS partial vertices performed in
ref.~\cite{quad}, where the dependence of $S$ and $D$ wave
vertices upon the relative energy is shown to be
smooth, contrary to the amplitudes which
display a strong dependence on $q_0$. Therefore, in our  calculations,
we can replace, at moderate values of $q_0$,
the $S$ and $D$ vertices by their values at $q_0=0$ with good accuracy.
The $P$ vertices can be expanded into Taylor series
around $q_0=0$ up to a desired order in $p_0/m$. Then the corresponding
derivatives can be computed numerically
along the imaginary axis since they are
analytical functions of $q_0$ \cite{anatomia}.

Finally, to cast our formulae in a more familiar form,
known from non-relativistic calculations,
we introduce the notion of BS wave functions
\cite{keisterTj,gross,quad}
\begin{eqnarray}
&&
\Psi_S(|{\bf P}_{lab}|) = {\cal N}
\frac{g_1(0,|{\bf P}_{lab}|)}{2E_p^{'}-M_D},
\quad
\Psi_D(|{\bf P}_{lab}|) = {\cal N}\frac{g_3(0,|{\bf P}_{lab}|)}{2E_2-M_D},
\label{positive}\\[2mm]
&&
\Psi_{P_5}(|{\bf P}_{lab}|) = {\cal N}\frac{g_5(0,|{\bf P}_{lab}|)}{M_D},
\quad
\Psi_{P_7}(|{\bf P}_{lab}|) = {\cal N}\frac{g_7(0,|{\bf P}_{lab}|)}{M_D},
\label{negative}
\end{eqnarray}
where ${\cal N} = 1/4\pi\sqrt{2M_D}$.
Then the cross section (\ref{crossBS}) and the amplitude (\ref{amplit})
may be computed in terms of
positive and negative-energy wave functions, $\Psi_{S,D}$
and $\Psi_{P_5,P_7}$ respectively.
The BS wave functions  $\Psi_{S,D}$ are
intimately related to the famous non-relativistic deuteron wave functions
$u(P_{lab}), w(P_{lab})$, and at small values of $P_{lab}$
they practically coincide~\cite{quad}. Therefore,
their contribution to the amplitudes and cross section is henceforth referred
to as the non-relativistic result.
The parts containing the negative-energy waves $\Psi_{P_5,P_7}$
are of a purely relativistic origin and consequently they manifest
genuine relativistic correction effects. The weights of these
wave functions are quite small~\cite{quad,rupp}
and we shall neglect all
the terms proportional to $\Psi_{P_5,P_7}^2$; only interferences
between $\Psi_{S,D}$ and $\Psi_{P_5,P_7}$  are kept in our
results. Besides the mentioned relativistic effects there
is another source of relativistic corrections, namely the so-called
Lorentz boost effects stemming from the transformation of
the BS wave functions from the c.m.s.\ to the deuteron rest frame.

\section{Lorentz boost effects} 

A straightforward way to study effects of the Lorentz boost
is to compare the non-relativistic results with
those obtained in the BS formalism by equating
to zero all contributions from the negative-energy waves leaving
only the contribution of the waves $\Psi_{S,D}$.
In the non-relativistic case, the cross section and the
spin amplitudes take a simple form \cite{rekalo,ladygin}
\begin{eqnarray}
&&
\frac{d\sigma_{NR}}{d\Omega}\,=\,3\,\left( u^2(q)+ w^2(q)\right)^2,
\label{crosrekal}
\\[2mm]
&&
{\cal A}_{NR}\,= \,\left( u(q)+\frac{w(q)}{\sqrt{2}}\right) ^2,\label{alad}\\
&&
{\cal B}_{NR}\,= \,-\frac{3}{2}w(q)\,\left( 2\sqrt{2}u(q)-w(q)\right ),
\label{blad}\\
 &&
{\cal C}_{NR}\,= \,\left( u(q)+\frac{w(q)}{\sqrt{2}}\right)
\left(u(q)-\sqrt{2}w(q)\right ),\label{clad}\\
&&
{\cal D}_{NR}\,=\,  \frac{3}{\sqrt{2}}w(q)\left( u(q)+\frac{w(q)}{\sqrt{2}}\right),
\label{dlad}
\end{eqnarray}
where in eqs.~(\ref{alad}) - (\ref{dlad}) the relative momentum
$q$ and appropriate sign conventions are
introduced in the definition of the
non-relativistic wave functions $u(q)$ and $ w(q)$.  Note that in
non-relativistic calculations there are ambiguities in treating
the internal momentum $q$. It is not clear
in what frame of reference
the  argument of the non-relativistic wave functions $u(q)$ and $ w(q)$
should be evaluated:
is $q$ to be computed in the
three-body center of mass frame as proposed in ref.~\cite{weber} or
should one use the deuteron rest frame \cite{nakamura} in defining the
deuteron wave function? At small momenta
these alternatives become identical, however a difference occurs already
at intermediate energies $P_{lab} \sim 0.2 \cdots$ 0.3 GeV/c.
In the covariant BS approach this problem is solved by
using invariant amplitudes $h_i(q^2,Dq)$ and by taking into account
the boost effects.

Substituting eqs.~(\ref{r1cm}) - (\ref{h16}) into  eqs.~(\ref{crossBS}) and
(\ref{observables})
and expanding the result into Taylor series
around $P_{lab}^2/2m^2$ and keeping the leading terms we obtain
the contribution of positive-energy BS waves in the form
\begin{eqnarray}
\frac{d\sigma_0}{d\Omega} & = & \frac{12m^2}{s}\,\left( \Psi_S^2(P_{lab})
+ \Psi^2_D(P_{lab})\right)^2\,P_{lab}^4
\left( 1+\frac{P_{lab}^2}{2m^2}+\frac{29P_{lab}^4}{16m^4}+
\frac{83P_{lab}^6}{32m^6} +\cdots\right ),
\label{crospositive}
\\[2mm]
{\cal A}_0 & = & 16\pi m P_{lab}^2\,\left(
\Psi_S(P_{lab})-\frac{\Psi_D(P_{lab})}{\sqrt{2}}\right) ^2
\,{\cal L}(P_{lab}),
\label{aBS}\\
{\cal B}_0 & = & 16\pi m P_{lab}^2\
\frac{3}{2}\Psi_D(P_{lab})\,\left( 2\sqrt{2}\Psi_S(P_{lab})
+ \Psi_D(P_{lab})\right )
\,{\cal L}(P_{lab}),
\label{bBS}\\
{\cal C}_0\,& = & 16\pi mP_{lab}^2\,\left( \Psi_S(P_{lab})
-\frac{\Psi_D(P_{lab})}{\sqrt{2}}\right)
\left(\Psi_S(P_{lab})+\sqrt{2}\Psi_D(P_{lab})\right )
\,{\cal L}(P_{lab}),
\label{cBS}\\
{\cal D}_0 & = & -\,16\pi mP_{lab}^2\,
\frac{3}{\sqrt{2}}\Psi_D(P_{lab})
\left( \Psi_S(P_{lab})-\frac{\Psi_D(P_{lab})}{\sqrt{2}}\right)\,
{\cal L}(P_{lab}),
\label{dBS}
\end{eqnarray}
where the Lorentz boost effects are represented by
$\,{\cal L}(P_{lab}) $ defined as
\begin{equation}
{\cal L}(P_{lab})=
\left( 1+\frac{P_{lab}^2}{4m^2}+\frac{7P_{lab}^4}{8m^4}
+\cdots \right ).
\label{lorentz}
\end{equation}
It is seen that the results within the BS
approach recover the non-relativistic formulae
in the leading order in $P_{lab}^2/2m^2$
and receive additional corrections from the Lorentz boost effects.
At small values of $P_{lab}$ these corrections are
negligible but at moderate values of $P_{lab}$  they become
important and may give up to 30 - 40\% contributions in the non-relativistic
cross section. Note that within the one-nucleon exchange mechanism,
$P_{lab}$ is kinematically restricted
so that the Taylor expansion in
eqs.~(\ref{crospositive}) - (\ref{lorentz})
is justified since $P_{lab}^2/2m^2 < 1$ in the whole
range of the initial energy $\sqrt{s}$.

The elastic cross section for a process with unpolarized
particles evaluated within the BS formalism
is shown in fig.~2. The dashed line is the contribution
of only positive-energy BS functions $\Psi_S$ and $\Psi_D$
(to be compared with the
dotted line which represents the computation of the
cross section in the non-relativistic limit with
Bonn potential \cite{Bonn_pot}).
The long-dashed curve represents the
corrections from pure Lorentz boost effects. The solid line is the result
of full BS calculations by exploiting the numerical solutions
\cite{parametrization,solution,quad} of the BS equation with a realistic
kernel with $\pi,\omega,\rho,\sigma,\eta,\delta$ exchanges.
Experimental data are taken from refs.~\cite{nakamura,dpelasexp}.
The relativistic effects coming
from the negative-energy waves are much smaller and
they are not displayed here. From fig.~2
it becomes clear that the Lorentz boost
effects become essential at
$P_{lab} >$ 0.5 GeV/c. This is an understandable effects,
since it is expected that the boost corrections
should increase with increasing initial energy.
Remind that the region $P_{lab} >$ 0.5 GeV/c corresponds already to
rather high initial energies, say $T_{kin}\sim$ 6 GeV.
(The kinematics of the process is so that
at $\sqrt{s}\to\infty$ the momentum of the
detected slow proton becomes $P_{lab} \to 0.75 \, m$.)

A comparison with experimental data shows that the
one-nucleon exchange mechanism alone does not describe
satisfactorily the cross section and that other mechanism should be
considered \cite{wilkin,nakamura}.
One should keep in mind, however, that
measurements of the cross section in the
strict backward direction are rather difficult and many experimental
data are presented as extrapolations of
$d\sigma/d\Omega$ obtained in nearly
backward direction to the exact backward angle
\cite{kislind,dpelasexp}. In view of the very strong angular dependence
of the cross section the data from different
groups differ noticeably
(see fig. 2 in ref. \cite{keisterTj} and
fig. 27 in ref. \cite{kislind}).
This, together with the above mentioned sensitivity of the results to the
chosen deuteron wave function,
generates uncertainties in a detailed comparisons with data.
It is not our aim here to improve the agreement with the data, but we
intend to proceed with our methodological study within the well defined
framework of the impulse approximation.

The Lorentz boost does not affect at all the
polarization characteristics defined by
eq.~(\ref{observables}), as it should be.  For instance,
a direct  calculation of  the cross section
eq.~(\ref{crossBS})  with the density matrix (\ref{rhocovariant})
results in the known non-relativistic formulae
for the tensor analyzing power
$T_{20} =-{\cal H}_{0,NN\to 0,0}/\sqrt{2}$
and the polarization transfer
$\kappa =3 {\cal H}_{0,L\to L,0} /2 $
\begin{eqnarray}
&&
T_{20}^{NR} = \frac{1}{\sqrt{2}}
\frac{-\Psi_D^2(P_{lab})-2\sqrt{2}\Psi_S(P_{lab})\Psi_D(P_{lab})  }
{\Psi_S^2(P_{lab})+ \Psi_D^2(P_{lab})},
\label{t20ner} \\[2mm]
&&
\kappa^{NR} = \frac{1}{\sqrt{2}}
\frac{\Psi_S^2(P_{lab}) -\Psi_D^2(P_{lab})
+\Psi_S(P_{lab})\Psi_D(P_{lab})/\sqrt{2}  }
{\Psi_S^2(P_{lab})+ \Psi_D^2(P_{lab})}.
\label{kappaner}
\end{eqnarray}
Observe that in the non-relativistic
limit the polarization characteristics eqs.~(\ref{t20ner})
and (\ref{kappaner}) in the elastic $pD$ elastic scattering
exactly coincide with ones in the reactions of inclusive
\cite{rekalo,experiment}
or exclusive \cite{cosy} deuteron break-up
processes. In the relativistic case such simple relations among
polarization observables do not hold. From this one may
conclude  that
a combined analysis of data obtained within different processes
would allow for an estimate of the role of relativistic corrections
in the deuteron wave function.

\section{Relativistic corrections} 

In this section we present results of numerical calculations of
the contribution to the polarization observables coming
from the negative-energy $P$ waves. We are going to
investigate the tensor analyzing power and polarization transfer.

There are two  factors of smallness in computing relativistic corrections:
the term $P_{lab}^2/2m^2$ and terms proportional to the negative-energy
$P$ waves in the BS amplitude. In estimating corrections
to eq.~(\ref {crospositive}) we expand the result
of the trace operation in eq.~(\ref{crossBS}) and
our results of calculating eqs.~(\ref{r1cm}) - (\ref{r6cm})
into Taylor series around $P_{lab}^2/2m^2$
and keep only the leading terms relative to negative-energy waves and
to $P_{lab}^2/2m^2$.
The final results for the relativistic  effects reads for the cross section
\begin{eqnarray}
&&
\frac{d\sigma}{d\Omega}\,=\,\frac{d\sigma_0}{d\Omega}\,+
\delta\sigma,
\label{sigmarelativ} \\[2mm]
&&
\delta\sigma\,=\,\frac{24\sqrt{6}m^3P_{lab}^3}{\sqrt{s}}
(\Psi_S^2+\Psi_D^2) \left ( \Psi_S+\sqrt{2} \Psi_D \right )
\left (
\Psi_{P_5}+ \frac{2\sqrt{2}P_{lab}^2}{3m^2}\Psi_{P_7}
\right )
\nonumber\\[2mm]
&& \hspace*{1cm}
+\frac{2\sqrt{6}mP_{lab}^5}{\sqrt{s}}
(\Psi_S^2+\Psi_D^2)
\left( 9\sqrt{2}\Psi_D+33\Psi_S\right)\Psi_{P_5}+\cdots
\nonumber
\end{eqnarray}
and for the tensor analyzing power $T_{20}$
\begin{eqnarray}
T_{20} & = & T_{20}^{NR} + \delta T_{20},
\label{t20rel} \\[2mm]
\hspace*{-1.5cm}
\delta T_{20} & = & -
\frac{2\sqrt{3} \left ( \Psi_S+\sqrt{2} \Psi_D \right )
\left (\Psi_S-\Psi_D/\sqrt{2}\right)^2}{(\Psi_S^2+\Psi_D^2)^2}
\frac{m}{P_{lab}}
\left (
\Psi_{P_5}+ \frac{2\sqrt{2}P_{lab}^2}{3m^2}\Psi_{P_7}+\cdots
\right )
\nonumber
\end{eqnarray}
and for the polarization transfer $\kappa$
\begin{eqnarray}
\kappa & = & \kappa^{NR} + \delta \kappa,
\label{kapparel}  \\[2mm]
\delta \kappa & = &
\frac{\sqrt{6}\left (\Psi_S-\Psi_D/\sqrt{2}\right)
\left[\left(2\sqrt{2}\Psi_D-\Psi_S\right )^2-9\Psi_D^2\right ]}
{(\Psi_S^2+\Psi_D^2)^2}
\frac{m}{P_{lab}}
\left (
\Psi_{P_5}+ \frac{2\sqrt{2}P_{lab}^2}{3m^2}\Psi_{P_7}+\cdots \right ).
\nonumber
\end{eqnarray}
It should be stressed that a comparison of
the relativistic corrections (\ref{t20rel}) and (\ref{kapparel})
with those obtained for the deuteron
break-up reaction \cite{semikh,break} demonstrates that
in the elastic proton-deuteron scattering
the polarization observables coincide with the ones
in the deuteron quasi-elastic and break-up processes
only in the non-relativistic limit.
Therefore, the data from these processes all
together (cf.~\cite{ls}) can determine
the magnitude of the relativistic
corrections and further constrain the deviation of the mechanism of these
processes from the simple one-nucleon exchange picture.

In fig.~3 results of calculations of the tensor analyzing power
$T_{20}$ in the elastic $pD$ backward reaction are presented.
The long-dashed curve is the contribution of only
positive-energy BS waves eq.~(\ref{t20ner}), the dotted line represents
the pure relativistic corrections in eq.~(\ref{t20rel}), while the solid
line is the total result within the BS formalism.
Experimental data (open and full circles) are from
refs.~\cite{dpelastic,ls,azh97}.
For the sake of completeness we present some experimental
data for the deuteron break-up processes \cite{experiment}
(triangles) and results of computation of $T_{20}$
within the minimal relativization scheme \cite{fs,relativz}
with the Paris deuteron wave function \cite{Paris}.
In fig.~3 it is seen that
the theoretical calculations predict a change of sign in
the $T_{20}$ whereas the experimental data from both deuteron
break-up and elastic scattering processes shows that $T_{20}$
remains negative in the whole interval of measured values of $P_{lab}$.
However, calculations with the positive-energy BS wave functions
(long-dashed line) together with  the
relativistic corrections  (dotted line) result in a better
description of data.

In fig.~4 similar calculations are presented  for
the polarization transfer $\kappa$. The dashed line is
the contribution of the positive-energy BS wave functions
eq.~(\ref{kappaner}), while relativistic corrections
eq.~(\ref{kapparel}) are presented
by the dotted line; the solid line is the total BS result.
As in the previous results the relativistic corrections
in $\kappa$ are small for small and moderate values of $P_{lab}$, but
become essential at higher values of the initial energies.

\section{Other spin observables} 

As mentioned in ref.~\cite{rekalo}, one of the goals of the
future experiments is a direct reconstruction of the four complex
amplitudes (\ref{amplit}). For this one needs to measure
7 independent observables, however as seen from
the above formulae, all polarization observables are
bilinear combinations of the amplitudes (\ref{amplit}) so
that the necessary number of measurements at given energy increases.
A full set of polarisation observables for complete
measurement has been proposed in refs.~\cite{rekalo,ladygin}.
It is found that 10 spin observables, i.e.
the  two of the first order, like cross section and tensor analyzing
power $ {\cal H}_{0,NN\to 0,0}$,  and
8 spin correlations of the second order, for instance
$ {\cal H}_{0,NN\to 0,NN}\,$, $ {\cal H}_{0,NN\to 0,SS}\,$,
$ {\cal H}_{0,N\to 0,LS}\,$, $ {\cal H}_{0,N\to N,0}\,$,
$ {\cal H}_{N,N\to 0,0}\,$, $ {\cal H}_{0,LS\to N,0}\,$,
$ {\cal H}_{0,LS\to 0,0}\,$ and ${\cal H}_{0,LN\to 0,LN}\,$
could provide a complete analysis of the spin amplitudes
(\ref{amplit}).

In the previous section the results of relativistic
calculations of the tensor analyzing power and
polarization transfer from the initial deuteron to the final proton
have been presented. In this section
additional calculations of spin-correlation observables
of the second order are performed. We consider here
the proton vector-vector transfer,
the deuteron vector-vector and tensor-tensor transfer coefficients.
They read explicitly
\begin{eqnarray}
&&
{\cal H}_{N,0\to N,0}\,\equiv \,
\frac{2\left( {3\cal A}^2 +2{\cal A}{\cal B} +{\cal B}^2 -2{\cal C}^2 -4{\cal C}{\cal D}-2{\cal D}^2\right )}
{ \mbox{Tr} \left ({\cal F}{\cal F}^+\right )}=
\frac{1}{9}
\frac{\left(\Psi_S+\sqrt{2}\Psi_D\right )^4 }
{\left(\Psi_S^2+\Psi_D^2\right )^2 }+
\label{tt1positive}\\[2mm]
&&
\frac{4\sqrt{6}}{9}
\frac{\left (\Psi_S-\Psi_D/\sqrt{2}\right )^2\left(\Psi_S+\sqrt{2}
\Psi_D\right )^3}{\left (\psi_S^2+\Psi_D^2\right )^3}
\frac{m}{P_{lab}} \left (
\Psi_{P_5}+ \frac{2\sqrt{2}P_{lab}^2}{3m^2}\Psi_{P_7}+\cdots
\right ), \nonumber
\end{eqnarray}
\begin{eqnarray}
&&
{\cal H}_{0,N\to 0,N}\,\equiv\, \frac{4\left( {\cal A}^2 +{\cal A}{\cal B} +{\cal C}^2\right )}
{ \mbox{Tr} \left ({\cal F}{\cal F}^+\right )}=
\frac{4}{9}\frac{\left ( \Psi_S-\Psi_D/\sqrt{2}\right )^2
\left (\Psi_S+\sqrt{2}\Psi_D\right )^2}{\left (\Psi_S^2+\Psi_D^2\right)^2}+
\label{tt2ner}\\[2mm]
&&\hspace*{-1.5cm}
\frac{4\sqrt{6}}{9}\frac{\left ( \Psi_S-\Psi_D/\sqrt{2}\right )^2
\left (\Psi_S+\sqrt{2}\Psi_D\right )  \left
( \Psi_S -2\sqrt{2}\Psi_D)^2-9\Psi_D^2\right )}
{\left (\psi_S^2+\Psi_D^2\right )^3}
\frac{m}{P_{lab}} \left (
\Psi_{P_5}+ \frac{2\sqrt{2}P_{lab}^2}{3m^2}\Psi_{P_7}+\cdots
\right ), \nonumber
\end{eqnarray}
\begin{eqnarray}
&&
{\cal H}_{0,NN\to 0,SS}\,\equiv\,
\frac{2\left(- 3{\cal A}^2 +2{\cal A}{\cal B} +3{\cal C}^2 +10{\cal C}{\cal D}+5{\cal D}^2\right )}
{ \mbox{Tr} \left ({\cal F}{\cal F}^+\right )}=
\frac{1}{4}\frac{\left (2\sqrt{2} \Psi_S+\Psi_D \right )^2
\Psi_d^2}{\left (\Psi_S^2+\Psi_D^2\right)^2}+
\label{vectorcorr}\\[2mm]
&&
\frac{\sqrt{6}\Psi_D\left (\Psi_S+\sqrt{2}\Psi_D\right )
\left( \Psi_S-\Psi_D/\sqrt{2}\right )^2\left(2\sqrt{2}\Psi_S+\Psi_D\right )}
{\left (\Psi_S^2+\Psi_D^2\right )^3}
\frac{m}{P_{lab}} \left (
\Psi_{P_5}+ \frac{2\sqrt{2}P_{lab}^2}{3m^2}\Psi_{P_7}+\cdots
\right ),\nonumber
\end{eqnarray}
where the first lines in each of eqs.~(\ref{tt1positive}) - (\ref{vectorcorr})
display the non-relativistic limit, while the second lines
are the corresponding parts of purely relativistic corrections.

Figs. 5 - 7 show these spin-correlation observables according to
eqs.~(\ref{tt1positive}) - (\ref{vectorcorr})
calculated within the BS formalism; they are depicted as solid lines.
The dashed lines show the contribution of the positive-energy
BS waves (i.e., the non-relativistic limit), while the dotted lines are
the relativistic corrections due to the contribution
of $P$ waves in the deuteron. It is seen that the
relativistic effects for the proton-proton transfer coefficients
are negligible (see fig.~5) while for the deuteron-deuteron
correlations these effects are essential at $P_{lab} \ge$ 0.5 GeV/c.

It is interesting to notice that there are
observables which in the non-relativistic
limit are exactly zero and therefore consist of relativistic corrections only.
For instance the tensor-tensor transfer coefficient
$ {\cal H}_{0,LN\to 0,LN}$ is predicted to vanish in the non-relativistic case,
while within the BS formalism one gets
\begin{eqnarray}
\mbox{Tr} \left ({\cal F}{\cal F}^+\right ) {\cal H}_{0,LN\to 0,LN} & = & 9\,
\left ( {\cal A}^2 + Re {\cal A}{\cal B} -{\cal C}^2\right )
\label{zerocorrelat} \\[2mm]
& \simeq &
54m^4P_{lab}^2 \left [
\sqrt{2}\Psi_S \left ( \Psi_{P_5}+2\sqrt{2}\Psi_{P_7}\right)
- \psi_D\left (7\Psi_{P_5}+2\sqrt{2}\Psi_{P_7}\right )\right ]^2,
\nonumber
\end{eqnarray}
where all the contributions from higher orders in
$P_{lab}^2/m^2$  have been neglected.
In spite of the small value of this tensor-tensor correlation
(see fig.~8)
the measurements of such observables may directly quantify the importance
of relativistic $P$ waves in the deuteron.

As a conclusion of this section we emphasize that
for some of the computed spin observables the relativistic
effects are not too important, whereas for a specific class
of observables (like the tensor-tensor correlations)
admixtures of $P$ waves in the deuteron may result in
important corrections, for instance the coefficient
(\ref{zerocorrelat})  differs
from zero only due to relativistic effects in the deuteron.

\section{One-iteration approximation} 

The relativistic corrections in
eqs.~(\ref{t20rel}) - (\ref{zerocorrelat})
are governed by negative-energy $P$ wave states in the deuteron.
This can be considered
as a hint that admixtures of $P$ waves within the BS approach
are related to relativistic corrections by
taking into account meson-exchange currents and N$\bar{\mbox{N}}$
pair production diagrams \cite{mec} in the non-relativistic picture.
To establish a correspondence between our results and  the
mentioned non-relativistic calculations we estimate
the contribution of the  relativistic corrections by
computing the $P$ wave vertices in the so-called
``one-iteration approximation'' \cite{karmanov}.
The gist of this approximation is as follows: In solving the
BS equation by an iteration procedure one puts as  zeroth
iteration the exact solution of the Schr\"odinger equation for
$S$ and $D$ vertices and zero for other waves; then the $P$
vertices are found by one iteration of the BS equation.
Our experience in solving numerically the BS equation shows that
it converges rapidly for relatively small momenta $< 1$ GeV/c. That means
when utilizing  the exact non-relativistic solutions,
after one iteration the resulting $P$ waves are not too far from the full
solution.

To obtain analytical expressions for the negative-energy waves we proceed with
the mixed BS equation in the sense that the BS vertices are expressed via the
BS amplitudes as
\begin{eqnarray}
&&
G(k) = i\int\,\frac{d^4p}{(2\pi)^4}\frac{\lambda^2_{mes}}{(p-k)^2-\mu^2}\,
\gamma_{mes}\,\Phi(p)\,\gamma_{mes} ,\label{bsmixed}
\end{eqnarray}
where $G(k)$  and $\Phi(p)$ are the BS vertex and amplitude respectively,
$\lambda^2_{mes}$ denotes the meson-nucleon coupling constant and
$\gamma_{mes} $ the meson-nucleon coupling vertex
(for scalar, pseudo-scalar or vector couplings).

Then using the decomposition of $G(k)$ and $\Phi(p)$ in the complete
set of spin angular matrices  $\Gamma_\alpha$ (for details consult
ref.~\cite{quad}) the BS
equation for partial vertices $g_\alpha $ and amplitudes $\Phi_\alpha$
(here $\alpha$ accounts for the $\rho$ spin indices of the
corresponding partial amplitude) may be written in the form
\begin{equation}
g_\alpha(k_0,|\vec k|) = i\int\,\frac{d^4p}{(2\pi)^4} d \Omega_k
\frac{\lambda^2_{mes}}{ (p-k)^2 - \mu^2 } \mbox{Tr}
\left [ \Gamma_\alpha^+(k_0,-\vec k)\gamma_{mes}
\Gamma_\beta (p_0,\vec p)\gamma_{mes}\right ]\Phi_\beta(p_0,|\vec p|).
\label{partial}
\end{equation}

Using the standard decomposition of the meson propagator over
generalized Legendre functions $Q_l(z)$,
where $z=(|\vec p\,|^2+ |\vec k|^2 +\mu^2 -(p_0-k_0)^2)/|\vec p\,||\vec k|$ ,
one gets
\begin{eqnarray}
&&
g_\alpha(k_0,|\vec k|) = -\lambda_{mes}^2\int
\frac{p^2 dp \, d p_0}{4\pi^2}
W_{\alpha\beta}(|\vec k|,|\vec p|) \,
\Phi_\beta(p_0,|\vec p|),
\label{wab} \\
&&
W_{\alpha\beta}( k, p)\equiv\, \sum_{lm}
\frac{Q_l(z)}{2\pi|\vec p||\vec k|}\,\int d\Omega_k d\Omega_p
Y_{lm}(p)Y^*_{lm}(k) \mbox{Tr}
\left [\Gamma_\alpha^+(k_0,-\vec k)\gamma_{mes}\Gamma_\beta(p_0,\vec p)
\gamma_{mes}\right ]. \nonumber
\end{eqnarray}
These expressions are still the exact BS equation in the ladder approximation.
Further we assume:\\
(i) in the first approximation the negative-energy waves
are exactly zero, i.e., in eqs.~(\ref{wab}) remain only
$\Phi_\beta = S^{++}$ and $ D^{++}$,\\
(ii) in the interaction kernel
$  W_{\alpha\beta}$  and in the vertex functions
$g^{++}(p)$ we neglect the dependence on
the relative energy, i.e.,
$W_{\alpha\beta}( k, p)\simeq W_{\alpha\beta}(|\vec k|,|\vec p|)$ and
$g^{++}(p) \simeq g^{++}(0,\vec p)$,\\
(iii) the negative-energy waves are obtained by only one iteration
of eq.~(\ref{wab}).

For the pseudo-scalar isovector exchange we get
\begin{eqnarray}
&&W_{P_1^{+-}\to S^{++}}=
W_{P_1^{-+}\to S^{++}}={\cal N} \left [ Q_0(z) |\vec k| - Q_1(z) |\vec p| \right ],\label{v51}\\
&&W_{P_1^{+-}\to D^{++}}=
W_{P_1^{-+}\to D^{++}}={\cal N} \sqrt{2} \left [ Q_2(z) |\vec k| - Q_1(z) |\vec p| \right ],\label{v52}\\
&&W_{P_3^{+-}\to S^{++}}=-
W_{P_3^{-+}\to S^{++}}={\cal N} \sqrt{2}\left [- Q_0(z) |\vec k| + Q_1(z) |\vec p| \right ],\label{v61}\\
&&W_{P_3^{+-}\to D^{++}}=-
W_{P_3^{-+}\to D^{++}}={\cal N} \left [- Q_1(z) |\vec p| + Q_2(z) |\vec k| \right ],\label{v521}
\end{eqnarray}
where ${\cal N} = -\sqrt{3}/(2|\vec p||\vec k|E_k)$,
$E_k = \sqrt{k^2+m^2}\sim m$.

We perform further calculations in the coordinate space:
\begin{eqnarray}
&&
\int \frac{p^2dp}{2|\vec p||\vec k|}\left [ |\vec k | Q_0(z) -|\vec p| Q_1(z)\right ]
\Psi_S(|\vec p|) = \int\, dr \frac{\Psi_S(r)}{r}{\rm e}^{ -\mu r} (1+\mu r)j_1(kr)\label{ur},\\[2mm]
&&
\int \frac{p^2dp}{2|\vec p||\vec k|}\left [ |\vec p | Q_1(z) -|\vec k| Q_2(z)\right ]
\Psi_D(|\vec p|) =- \int\, dr \frac{\Psi_D(r)}{r}{\rm e}^{ -\mu r} (1+\mu r)j_1(kr)\label{wr},
\end{eqnarray}
where  $\Psi_S(r)$ and $\Psi_D(r)$ are the deuteron wave functions in the
coordinate space.

Then  the result for
the function $\Psi_{P_{5,7}}$  with a
BS kernel with pseudo-scalar one-boson exchange reads
\begin{equation}
\Psi_{P_{5,7}}(P_{lab}  ) = -g_\pi^2
\frac{2\sqrt{3}}{M_dE_p'}
\int\limits_0^\infty dr \, \frac{{\rm e}^{-\mu r}}{r} \, (1+\mu r) \,
{\rm j_1}(r P_{lab})
\left [
N_u \, u(r) + N_w \, w(r)\right ],\label{uwr}
\end{equation}
where $u(r)$ and $w(r)$ are the non-relativistic deuteron wave functions
in the coordinate representation, and $g_\pi^2 \approx$ 14.5
is the pion-nucleon coupling constant.
The normalization factors are
$N_u = \sqrt{2}$ (1) and $N_w =$  -1 ($\sqrt{2}$) for
$\Psi_{P_5}$ $(\Psi_{P_7})$ waves.

With this definition of the negative-energy waves one may estimate
the origin of the relativistic corrections computed
within the non-relativistic limit as additional contribution to the impulse
approximation diagrams, such as meson exchange currents and
$N\bar N $ pair production currents. As an example
we compute within the one-iteration approximation
the amplitude $\cal A$ which turns out to have
the simple form of a negative-energy wave contribution
\begin{equation}
{\cal A} = {\cal A}_0\, +\,32\sqrt{6}\pi P_{lab}^3
\left(\Psi_S- \frac{1}{\sqrt{2}} \Psi_D\right ) \Psi_{P_5}.
\label{Aoneit}
\end{equation}
Substituting (\ref{uwr})  into the expression
for the amplitude eq.~(\ref{Aoneit}), the relativistic corrections
in the one-iteration approximation become
\begin{eqnarray} \hspace*{-1cm}
\delta {\cal A} &=& -g_\pi^2 P_{lab}^3
\frac{192 \sqrt{2}\pi}{M_dE_p'} \nonumber \\
&\times&
\int\limits_0^\infty dr \, \frac{{\rm e}^{-\mu r}}{r} \, (1+\mu r) \,
{\rm j_1}(r P_{lab})
\left [
\sqrt{2} \, u(r) - \, w(r)\right ]
\left(\Psi_S- \frac{1}{\sqrt{2}} \Psi_D\right ),
\label{dp14}
\end{eqnarray}
which is similar to expressions obtained in non-relativistic evaluations
of the so-called ``catastrophic'' and pair production   diagrams
in electro-disintegration processes of
the deuteron \cite{foldy} and which is also similar to
results of computation of the triangle diagrams usually considered
in the elastic $pD$ processes \cite{wilkin,nakamura}. In our
case these corrections may be represented as diagrams with
meson exchange due to anti-nucleon degrees of freedom in
the BS equation, as depicted in fig.~9.
The remaining amplitudes
${\cal B}, {\cal C}, {\cal D}$ and, consequently, the cross section
(\ref{sigmarelativ}) and all the polarization observables (\ref{observables})
receive analogous corrections. From this it becomes clear that
generic relativistic calculations, even in impulse approximation,
contain already to some extent specific meson-exchange diagrams, i.e.,
pair production currents, and one should pay attention
on the problem of double counting when computing relativistic corrections
beyond the spectator mechanism.

\section{Summary} 

In summary, we present an explicit analysis of various relativistic effects
in  elastic backward scattering of protons off deuterons
within the Bethe-Salpeter
formalism with a realistic interaction kernel.
To have a well defined framework for our methodological investigations
we rely here on the impulse approximation. This allows to identify and
investigate separately
the contributions of the positive-energy
waves, Lorentz boost corrections and relativistic effects
due to negative-energy waves. Particular
attention is paid to the computation of the four
spin amplitudes of the process within the Bethe-Salpeter
approach. By writing these amplitudes in the center of mass
system in a non-covariant form a direct correspondence
between our approach
and the general phenomenological analysis of the process is
found. In such a way a suitable representation of
the polarization observables and a straightforward
investigation of the non-relativistic limit are achieved.

Numerical estimates of the Lorentz boost and
other relativistic effects in the cross section and selected
polarization observables, at kinematical conditions of ongoing and
forthcoming experiments \cite{proposal,cosy} are presented.
It is found that in a complete set of polarization
observables, proposed for a reconstruction of the amplitude,
relativistic corrections either may be negligible for a certain class of
spin-correlations or play a crucial role for other
observables.
It is shown that the one-nucleon exchange mechanism alone
does not give the predominant contribution in these reactions
and future experiments
must clarify effects beyond the impulse approximation.

\subsubsection*{Acknowledgments} 

Useful discussions with
A.Yu. Umnikov and F. Santos are gratefully acknowledged.
Two of the authors (L.P.K. and S.S.S.) would like
to thank for the warm hospitality of the nuclear theory group
at the Research Center Rossendorf.
This work has been supported in parts by a grant
of the Heisenberg-Landau JINR-FRG collaboration project.

\section*{Appendix A} 

\def\theequation{A\,\arabic{equation}}
\setcounter{equation}{0}

\noindent
The covariant expression of the vertex function
$\Gamma(D,q)$ for the BS equation
with one particle on mass shell is rather known and
may be found, for instance in refs.~\cite{keisterTj,gross,rupp}.
Here we present the covariant solution $\Gamma(D,q)$
for the full BS equation when both particles are off the mass shell
\begin{eqnarray}
\Gamma(D,q) = [h_1 \hat{\xi} +h_2 \frac {(q \xi)}{m}] +
\frac {\hat{D}/2+\hat{q}-m}{m} [h_3 \hat{\xi} +h_4 \frac {(q \xi)}{m}] +
\nonumber \\[0mm]
[h_5 \hat{\xi} +h_6 \frac {(q \xi)}{m}] \frac {\hat{D}/2-\hat{q}+m}{m}+
\frac {\hat{D}/2+\hat{q}-m}{m}
[h_7 \hat{\xi} +h_8 \frac {(q \xi)}{m}] \frac {\hat{D}/2-\hat{q}+m}{m}.
\end{eqnarray}
The eight invariant
scalars $h_i(Dq,q^2)$ are  connected with the corresponding spin-orbit
momentum vertex functions $g_i(p_0,P_{lab})$,
which are numerically determined in the deuteron rest frame, via
\begin{eqnarray}
\sqrt{4\pi}h_1 & = & \frac{\sqrt{2}}{16E_p^{'}M_d}
(2E_p^{'}-2p_{0}+M_d)(M_d+2p_{0}+2E_p^{'}) g_1 +\nonumber\\
&&
\frac{\sqrt{2}}{16E_p^{'}M_d} (-M_d+2p_{0}+2E_p^{'})(2E_p^{'}-2p_{0}-M_d) g_2 -
\nonumber \\
&&
\frac{1}{16E_p^{'}M_d} (2E_p^{'}-2p_{0}+M_d)(M_d+2p_{0}+2E_p^{'}) g_3 -
\nonumber\\
&&
\frac{1}{16E_p^{'}M_d} (-M_d+2p_{0}+2E_p^{'})(2E_p^{'}-2p_{0}-M_d) g_4 +
\nonumber\\
&&
+ \frac{\sqrt{3}m}{16P_{lab}M_dE_p^{'}} (-M_d+2p_{0}+2E_p^{'})(M_d+2p_{0}+2E_p^{'}) g_5 -
\nonumber\\
&&
\frac{\sqrt{3}m}{16P_{lab}M_dE_p^{'}} (2E_p^{'}-2p_{0}-M_d)(2E_p^{'}-2p_{0}+M_d) g_6,
\label{h1} \\
\sqrt{4\pi}h_2 & = &
- \frac{\sqrt{2}m}{16(E_p^{'}+m)M_dE_p^{'}} (2E_p^{'}-2p_{0}+M_d)(M_d+2p_{0}+2E_p^{'}) g_1-
\nonumber \\
&&
\frac{\sqrt{2}m}{16(E_p^{'}+m)M_dE_p^{'}} (-M_d+2p_{0}+2E_p^{'})(2E_p^{'}-2p_{0}-M_d) g_2 -
\nonumber \\
&&
\frac{(m+2E_p^{'})m}{16(E_p^{'}-m)(E_p^{'}+m)E_p^{'}M_d} (2E_p^{'}-2p_{0}+M_d)(M_d+2p_{0}+2E_p^{'}) g_3 -
\nonumber \\
&&
\frac{(m+2E_p^{'})m}{16(E_p^{'}-m)(E_p^{'}+m)E_p^{'}M_d} (-M_d+2p_{0}+2E_p^{'})(2E_p^{'}-2p_{0}-M_d) g_4 -
\nonumber\\
&&
\frac{\sqrt{3}m}{16pM_dE_p^{'}} (-M_d +2p_{0}+2E_p^{'})(M_d+2p_{0}+2E_p^{'}) g_5 +
\nonumber\\
&&
\frac{\sqrt{3}m}{16pM_dE_p^{'}} (2E_p^{'}-2p_{0}-M_d)(2E_p^{'}-2p_{0}+M_d) g_6,
\label{h2} \\
\sqrt{4\pi}h_3 & = & \frac{\sqrt{3}m}{8M_dP_{lab}} (-M_d+2p_{0}+2E_p^{'}) g_5-
\nonumber\\
&&
\frac{\sqrt{3}m}{8M_dP_{lab}} (2E_p^{'}-2p_{0}+M_d) g_6,
\label{h3} \\
\sqrt{4\pi}h_4 & = &
\frac{\sqrt{2}m^{2}}{8(E_p^{'}+m)M_dE_p^{'}} (2E_p^{'}-2p_{0}+M_d) g_1+
\nonumber\\
&&
\frac{\sqrt{2}m^{2}}{8(E_p^{'}+m)M_dE_p^{'}} (-M_d+2p_{0}+2E_p^{'}) g_2-
\nonumber \\
&&
\frac{(E_p^{'}+2m)m^{2}}{8(E_p^{'}-m)(E_p^{'}+m)E_p^{'}M_d} (2E_p^{'}-2p_{0}+M_d) g_3-
\nonumber\\
&&
\frac{(E_p^{'}+2m)m^2}{8(E_p^{'}-m)(E_p^{'}+m)E_p^{'}M_d} (-M_d+2p_{0}+2E_p^{'}) g_4+
\nonumber \\
&&
\frac{\sqrt{3}\sqrt{2}m^2}{8M_dP_{lab}E_p^{'}} (-M_d+2p_{0}+2E_p^{'}) g_7-
\nonumber\\
&&
\frac{\sqrt{3}\sqrt{2}m^2}{8M_dP_{lab}E_p^{'}} (2E_p^{'}-2p_{0}+M_d)g_8,
\label{h4} \\
\sqrt{4\pi}h_5 & = &
- \frac{m\sqrt{3}}{8M_dP_{lab}} (M_d+2p_{0}+2E_p^{'}) g_5+
\nonumber\\
&&
\frac{m\sqrt{3}}{8M_dP_{lab}} (2E_p^{'}-2p_{0}-M_d) g_6,
\label{h5} \\
\sqrt{4\pi}h_6 & = &
- \frac{m^{2}\sqrt{2}}{8(E_p^{'}+m)M_dE_p^{'}} (M_d+2p_{0}+2E_p^{'}) g_1-
\nonumber\\
&&
\frac{m^{2}\sqrt{2}}{8(E_p^{'}+m)M_dE_p^{'}} (2E_p^{'}-2p_{0}-M_d) g_2+
\nonumber \\
&&
\frac{(E_p^{'}+2m)m^{2}}{8(E_p^{'}-m)(E_p^{'}+m)E_p^{'}M_d} (M_d+2p_{0}+2E_p^{'}) g_3+
\nonumber\\
&&
\frac{(E_p^{'}+2m)m^{2}}{8(E_p^{'}-m)(E_p^{'}+m)E_p^{'}M_d} (2E_p^{'}-2p_{0}-M_d) g_4+
\nonumber \\
&&
\frac{\sqrt{3}\sqrt{2}m^{2}}{8M_dP_{lab}E_p^{'}} (M_d+2p_{0}+2E_p^{'}) g_7-
\nonumber\\
&&
\frac{\sqrt{3}\sqrt{2}m^{2}}{8M_dP_{lab}E_p^{'}} (2E_p^{'}-2p_{0}-M_d) g_8,
\label{h6} \\
\sqrt{4\pi}h_7 & = & \frac{\sqrt{2}m^{2}}{4M_dE_p^{'}} g_1
+ \frac{\sqrt{2}m^{2}}{4M_dE_p^{'}} g_2
- \frac{m^{2}}{4M_dE_p^{'}} g_3
\nonumber\\
&&
- \frac{m^{2}}{4M_dE_p^{'}} g_4
- \frac{m^{3}\sqrt{3}}{4M_dP_{lab}E_p^{'}} g_5
+ \frac{m^{3}\sqrt{3}}{4M_dP_{lab}E_p^{'}} g_6,
\label{h7} \\
\sqrt{4\pi}h_8 & = &
\frac{\sqrt{2}m^{3}}{4(E_p^{'}+m)M_dE_p^{'}} g_1
+ \frac{\sqrt{2}m^{3}}{4(E_p^{'}+m)M_dE_p^{'}} g_2+
\nonumber\\
&&
\frac{m^{3}(m+2E_p^{'})}{4(E_p^{'}-m)(E_p^{'}+m)E_p^{'}M_d} g_3
+ \frac{m^{3}(m+2E_p^{'})}{4(E_p^{'}-m)(E_p^{'}+m)E_p^{'}M_d} g_4
\nonumber \\
&&
- \frac{m^{3}\sqrt{3}}{4M_dP_{lab}E_p^{'}} g_5
+ \frac{m^{3}\sqrt{3}}{4M_dP_{lab}E_p^{'}} g_6,
\label{h8}
\end{eqnarray}
where $p_0=(Dq)/M_d$ and $E_p^{'} =\sqrt{P_{lab}^2+m^2}$.
The spin-orbit momentum
vertex functions are defined within the $\rho$ spin classification as
$g_1 = S^{++}$, $g_2 = S^{--}$,
$g_3 = D^{++}$, $g_4 = D^{--}$, $g_5 = ^3\!\!P^{+-}$, $g_6 = ^3\!\!P^{-+}$,
$g_7 = ^1\!\!P^{+-}$, $g_8 = ^1\!\!P^{-+}$.

Eqs.~(\ref{h1}) - (\ref{h8}) may be written in a more
compact form, however the present expressions
are more informative and easily understood by the reader.

\section{Appendix B} 

\def\theequation{B\,\arabic{equation}}
\setcounter{equation}{0}

\noindent
In obtaining eqs.~(\ref{r1cm}) - (\ref{r6cm})
the following relations are useful
\begin{eqnarray}
&& \hspace*{-2cm}
p \xi'=\frac {E-\epsilon}{M_d}(\mbox{\boldmath{$p$}} \mbox{\boldmath{$\xi$}}'), \quad
p' \xi=-\frac {E-\epsilon}{M_d}(\mbox{\boldmath{$p$}} \mbox{\boldmath{$\xi$}}),
\\[4mm]
&& \hspace*{-2cm}
\big ([{\bf a}\times {\bf b}]\big)
\big ([{\bf c}\times {\bf d}]\big)=
({\bf a}{\bf c}) ({\bf b}{\bf d}) -
({\bf a}{\bf d}) ({\bf b}{\bf c}),
\\[4mm]
&&\hspace*{-2cm}
-i \mbox{\boldmath{$\sigma$}} ( \mbox{\boldmath{$p$}} \times \mbox{\boldmath{$\xi$}})(\mbox{\boldmath{$p$}} \mbox{\boldmath{$\xi$}}')
+i \mbox{\boldmath{$\sigma$}} ( \mbox{\boldmath{$p$}} \times \mbox{\boldmath{$\xi$}}')(\mbox{\boldmath{$p$}} \mbox{\boldmath{$\xi$}}) =
i \mbox{\boldmath{$\sigma$}}( \mbox{\boldmath{$\xi$}} \times \mbox{\boldmath{$\xi$}}')p^2
-i (\mbox{\boldmath{$\sigma$}} \mbox{\boldmath{$p$}}) ( \mbox{\boldmath{$p$}}, \mbox{\boldmath{$\xi$}} \times \mbox{\boldmath{$\xi$}}'),
\\[4mm]
&&
\hspace*{-2cm}
\mbox{\boldmath{$a$}}=\mbox{\boldmath{$\xi$}}+\mbox{\boldmath{$p$}} \frac {\mbox{\boldmath{$p$}} \mbox{\boldmath{$\xi$}}}{M_d(E+M_d)}, \quad
\mbox{\boldmath{$a$}}'=\mbox{\boldmath{$\xi$}}'+\mbox{\boldmath{$p$}} \frac {\mbox{\boldmath{$p$}} \mbox{\boldmath{$\xi$}}'}{M_d(E+M_d)},
\\[4mm]
&&
\hspace*{-2cm}
(\mbox{\boldmath{$\sigma$}} \mbox{\boldmath{$a$}}) (\mbox{\boldmath{$\sigma$}} \mbox{\boldmath{$a$}}')= (\mbox{\boldmath{$\xi$}} \mbox{\boldmath{$\xi$}}') +
i \mbox{\boldmath{$\sigma$}}( \mbox{\boldmath{$\xi$}} \times \mbox{\boldmath{$\xi$}}') \frac{E}{M_d}+
\frac {\mbox{\boldmath{$p$}} \mbox{\boldmath{$\xi$}}}{M_d} \frac {\mbox{\boldmath{$p$}} \mbox{\boldmath{$\xi$}}'}{M_d}
-i (\mbox{\boldmath{$\sigma$}} \mbox{\boldmath{$p$}}) ( \mbox{\boldmath{$p$}}, \mbox{\boldmath{$\xi$}} \times \mbox{\boldmath{$\xi$}}') \frac {1}{M_d(M_d+E)}.
\end{eqnarray}

\newpage


\begin{figure}[h]
\centerline{\epsfxsize=.6 \hsize \epsffile{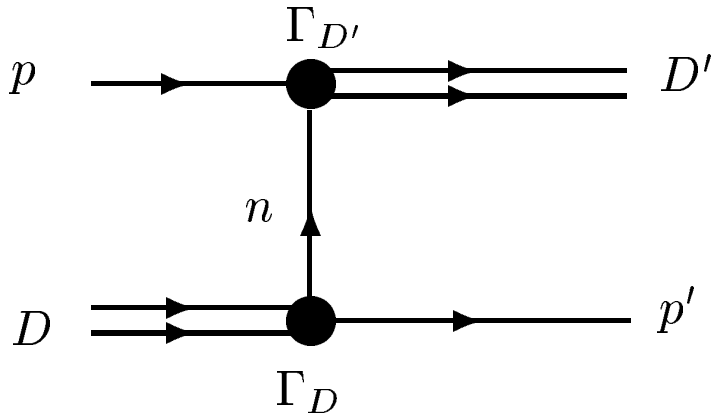}}
\vskip 0.6cm
\caption{
The one-nucleon exchange graph for the reaction
$p +\, D\, =\, p'(\Theta=180^o)\,+\,D'$.}
\end{figure}

\begin{figure}
\centerline{\epsfxsize=.7 \hsize \epsffile{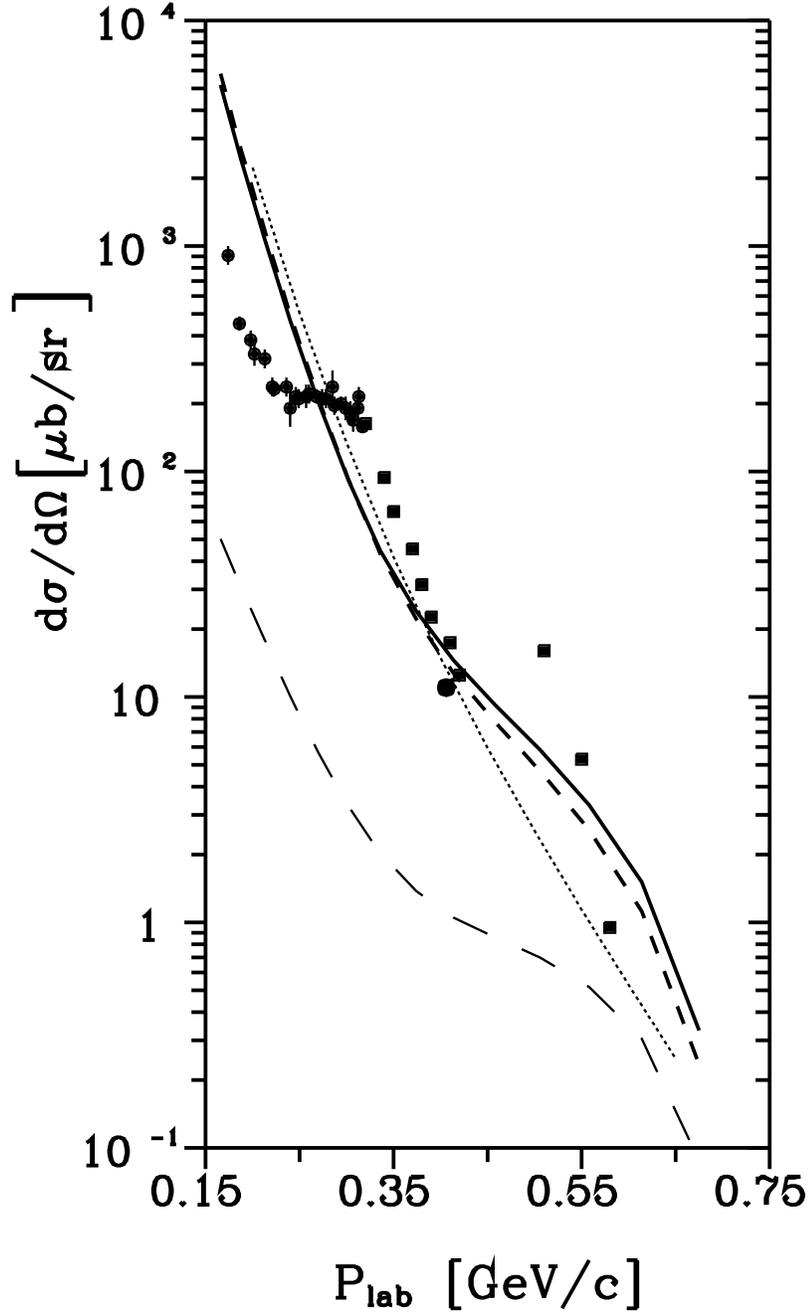}}
\caption{
The spin averaged differential cross section
$d\sigma/d\Omega$
for the elastic proton-deuteron backward scattering
in the c.m.s. as a function
of the momentum of the detected proton in the laboratory system.
Dashed line: contribution of the positive-energy BS waves,
long-dashed line: contribution of the Lorentz-boost effects
eq.~(\protect\ref{lorentz}),
solid line: full BS calculations,
dotted line: results of calculations within the
non-relativistic limit with the Bonn potential wave function.
Experimental data  from \protect\cite{nakamura,dpelasexp}.}
\end{figure}

\begin{figure}
\centerline{\epsfxsize=.9 \hsize \epsffile{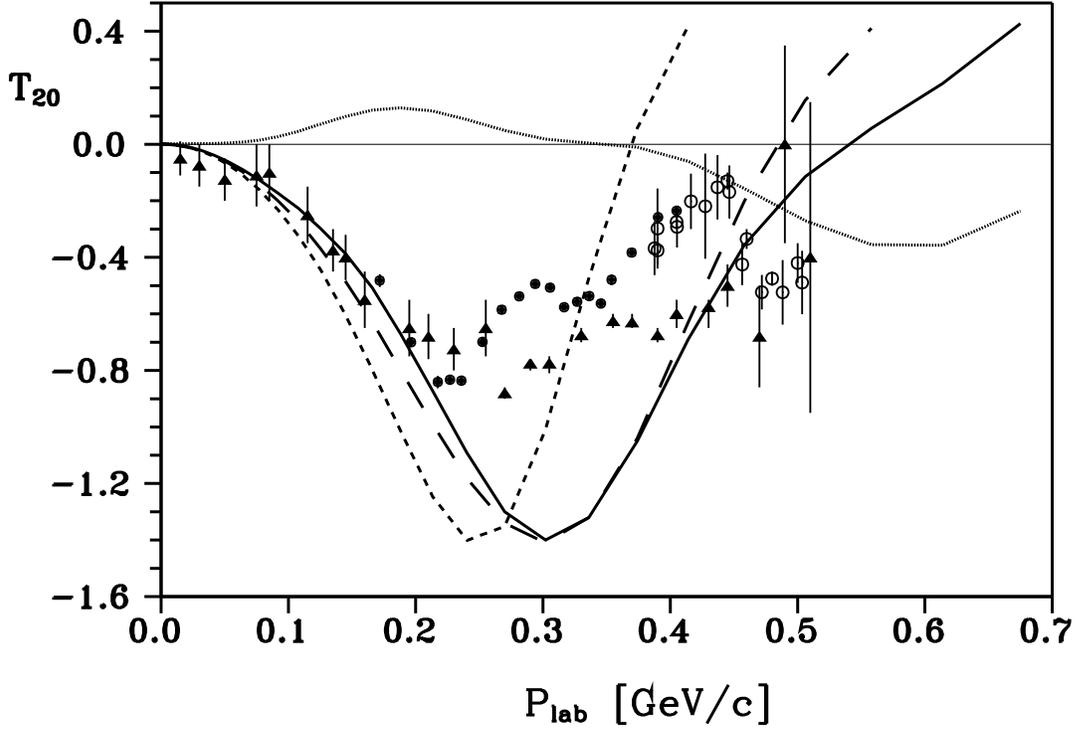}}
\caption{
The deuteron tensor analyzing power $T_{20}$
for the elastic proton-deuteron backward scattering.
Long-dashed line: contribution of the positive
energy BS waves eq.~(\protect\ref{t20ner}),
dotted line: purely relativistic corrections computed by
eq.~(\protect\ref{t20rel}),
solid line: results of computation within the BS approach
eq.~(\protect\ref{t20rel}),
short-dashed line: results of computation
within the minimal relativization scheme \protect\cite{relativz}
with Paris potential wave function.
Experimental data: circles - elastic
backward scattering \protect\cite{dpelastic,punj95,ls,azh97},
triangles -  $T_{20}$ measured in the deuteron
break-up reaction \protect\cite{experiment}.}
\end{figure}

\begin{figure}
\centerline{\epsfxsize=.7 \hsize \epsffile{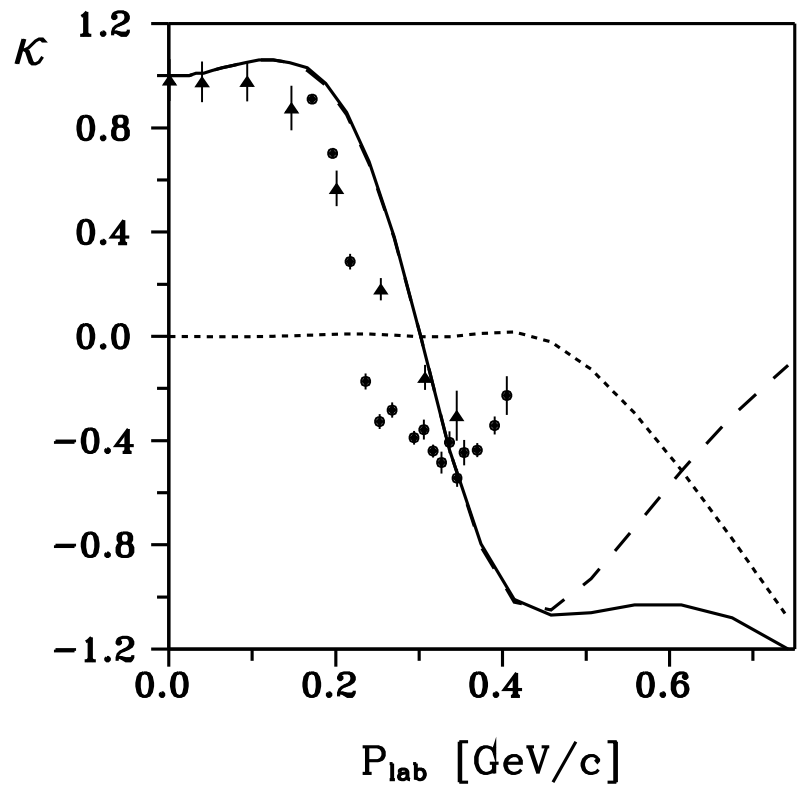}}
\caption{
The polarization transfer $\kappa$
for the elastic proton-deuteron backward scattering.
Notation as in fig.~3.}
\end{figure}

\begin{figure}
\centerline{\epsfxsize=.9 \hsize \epsffile{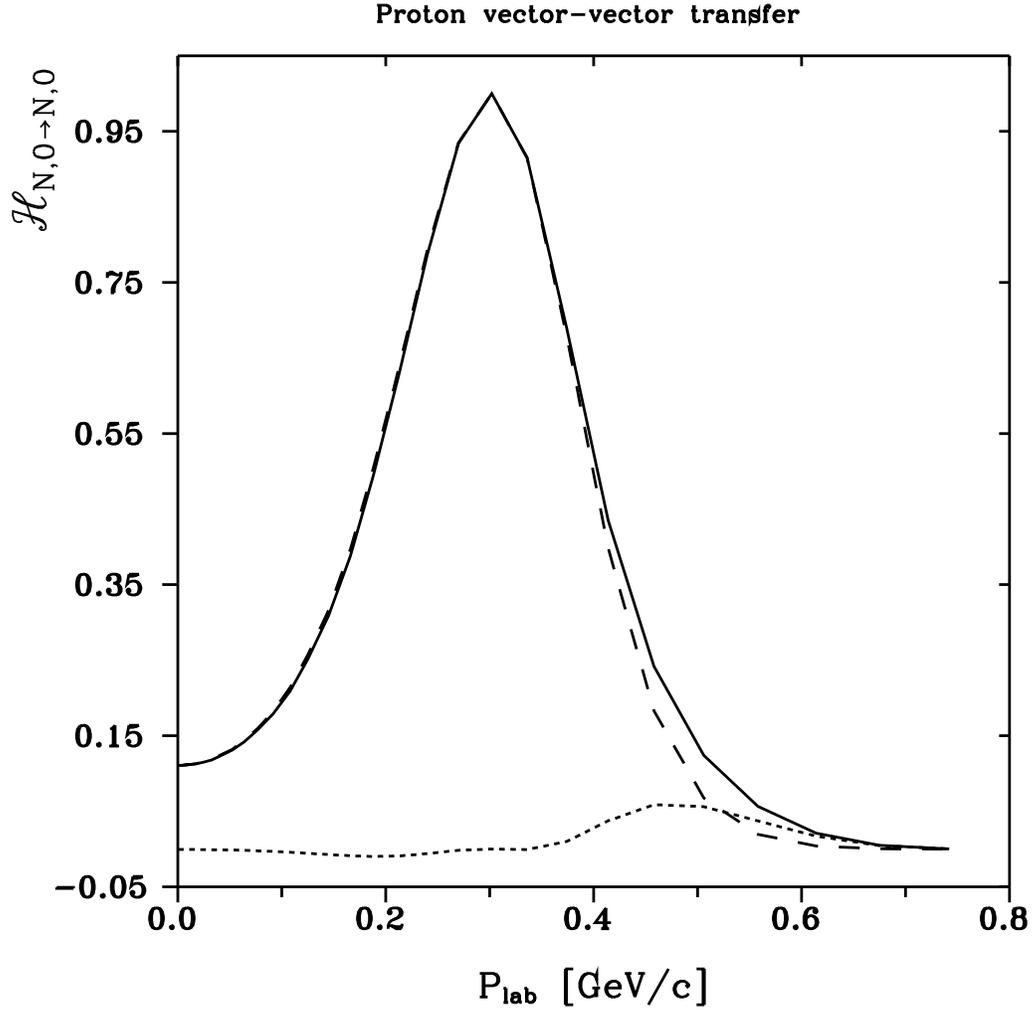}}
\caption{
The vector-vector polarization transfer coefficient
from the initial proton to the final proton.
Dashed line: contribution of the positive-energy BS
waves (i.e., the non-relativistic limit),
dotted line: relativistic corrections,
solid line: full BS results via eq.~(\protect\ref{tt1positive}).}
\end{figure}

\begin{figure}
\centerline{\epsfxsize=.9 \hsize \epsffile{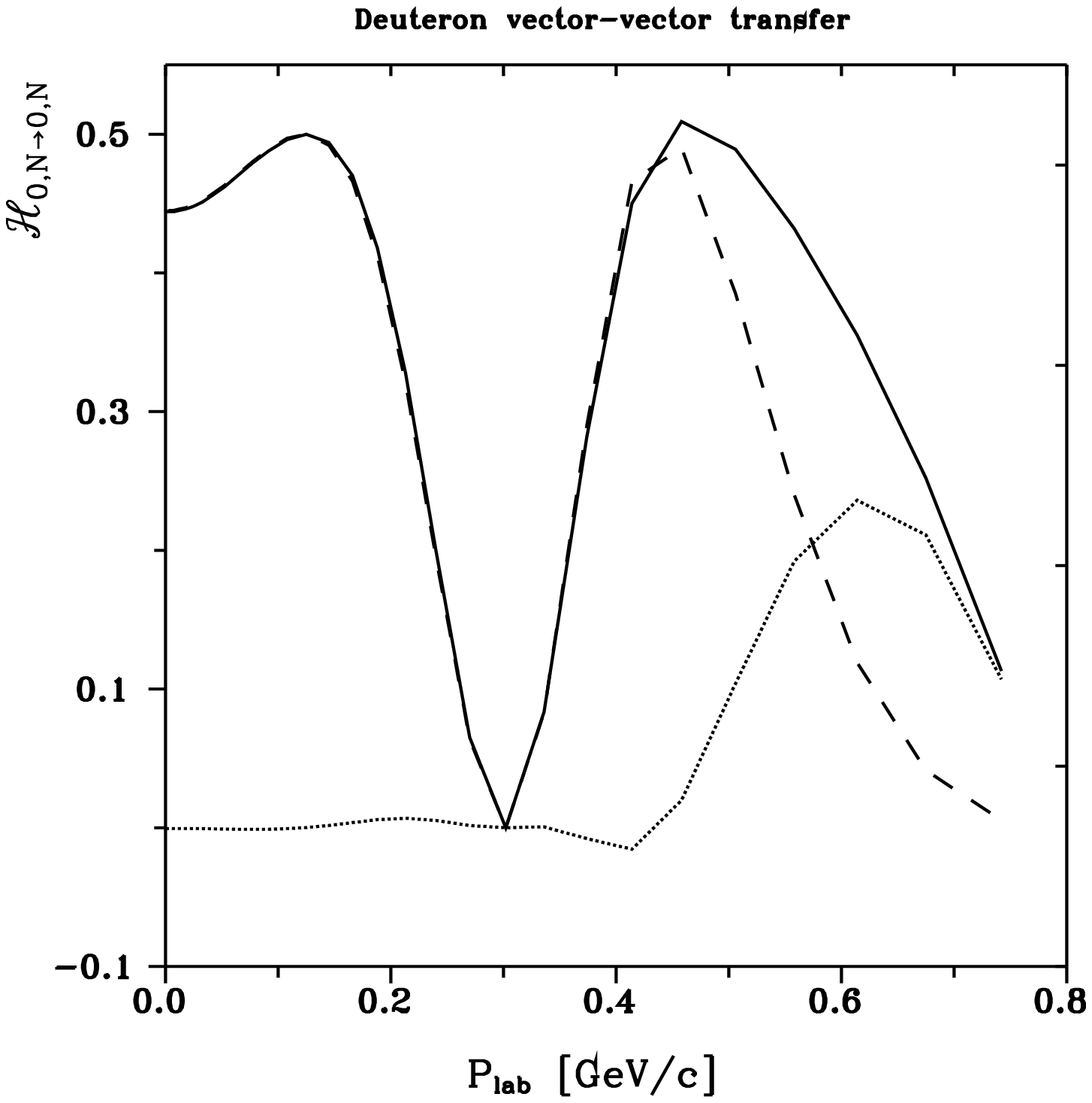}}
\caption{
The vector-vector polarization transfer coefficient
from the initial deuteron to the final deuteron.
Notation as in fig.~5}
\end{figure}

\begin{figure}
\centerline{\epsfxsize=.9 \hsize \epsffile{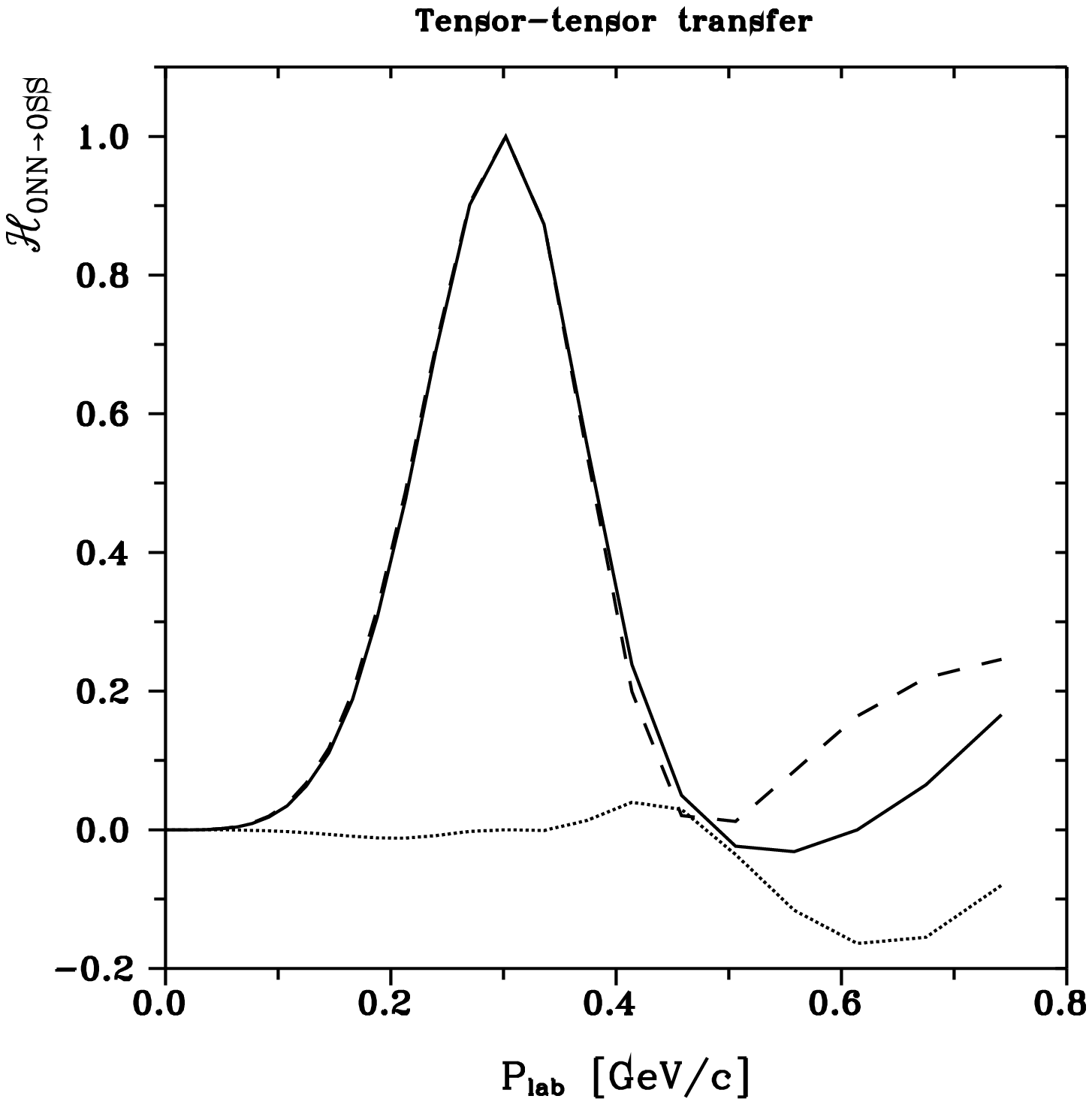}}
\caption{
The tensor-tensor polarization transfer coefficient
from the initial deuteron to the final deuteron.
Notation as in fig.~5.}
\end{figure}

\begin{figure}
\centerline{\epsfxsize=.7 \hsize \epsffile{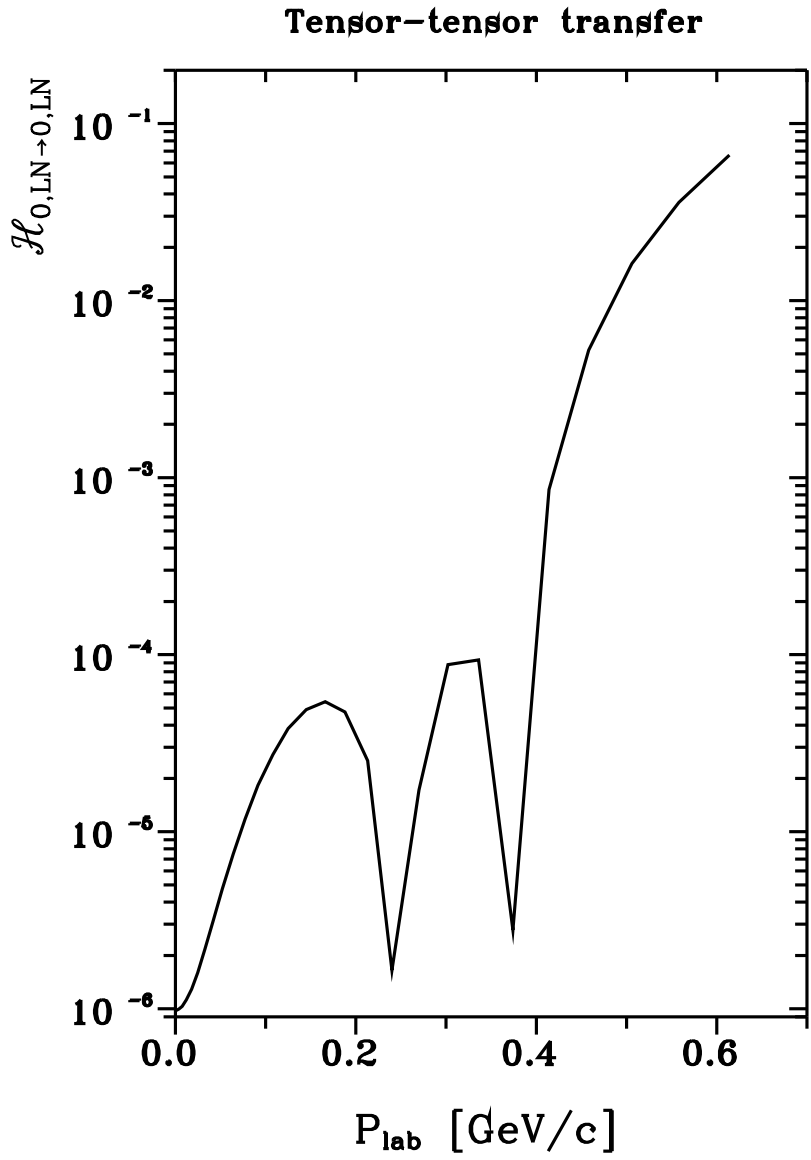}}
\caption{
The relativistic corrections for the
tensor-tensor polarization transfer coefficient
from the initial deuteron to the final deuteron
$ {\cal H}_{0,LN\to 0,LN}$ defined by eq.~(\protect\ref{zerocorrelat}).
In the non-relativistic limit $ {\cal H}_{0,LN\to 0,LN}$ vanishes.}
\end{figure}

\begin{figure}
\centerline{\epsfxsize=.7 \hsize \epsffile{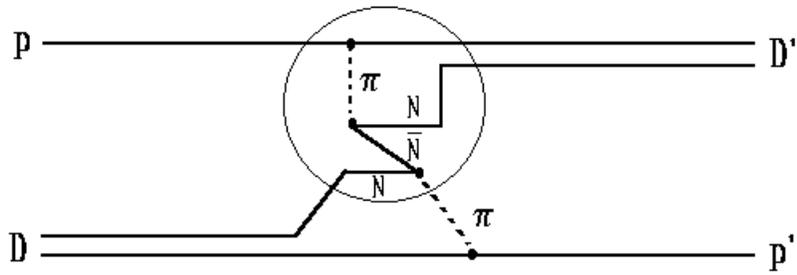}}
\caption{
A possible intermediate mechanism already included in the
one-nucleon exchange diagram in the BS approach.}
\end{figure}

\end{document}